\documentclass[journal=jpcck,manuscript=article]{achemso}

\usepackage{amsmath,amssymb,graphicx}% Include figure files
\usepackage{dcolumn}% Align table columns on decimal point
\usepackage{bm}% bold math 
\usepackage[usenames]{color}

\newcommand{\nG}{n_\textrm{G}}
\newcolumntype{L}{>{$}l<{$}} % math-mode version of "l" column type 

\author{Isabelle M. Palstra}
\affiliation{Institute of Physics, University of Amsterdam, Science Park 904, 1098 XH Amsterdam, The Netherlands}
\alsoaffiliation{Center for Nanophotonics, AMOLF, Science Park 104, 1098 XG, Amsterdam, The Netherlands}
\author{A. Femius Koenderink}
\affiliation{Center for Nanophotonics, AMOLF, Science Park 104, 1098 XG, Amsterdam, The Netherlands}
\email{f.koenderink@amolf.nl}

\title{A Python toolbox for unbiased statistical analysis of fluorescence intermittency of multi-level emitters}


\date{\today}

\begin{document}
\begin{abstract}
We report on a Python-toolbox for unbiased statistical analysis of fluorescence intermittency properties of single emitters. Intermittency, i.e., step-wise temporal variations in the instantaneous emission intensity and fluorescence decay rate properties are common to organic fluorophores, II-VI quantum dots and perovskite quantum dots alike. Unbiased statistical analysis of intermittency switching time distributions, involved levels and lifetimes is important to avoid interpretation artefacts. This work provides an implementation of Bayesian changepoint analysis and level clustering applicable to time-tagged single-photon detection data of single emitters that can be applied to real experimental data and as tool to verify the ramifications of hypothesized mechanistic intermittency models. We provide a detailed Monte Carlo analysis to illustrate these statistics tools, and to benchmark the extent to which conclusions can be drawn on the photophysics of highly complex systems, such as perovskite quantum dots that switch between a plethora of states instead of just two. 
\end{abstract}

\maketitle 
 
\section{Introduction}
Since the seminal first observation of single molecule emitters  in fluorescence microscopy three decades ago~\cite{Orrit1990}, single quantum emitter photophysics has taken center stage in a large body of research. On one hand, single quantum emitters as single photon sources~\cite{Lounis2005} are held to be an essential part of quantum communication networks  and are deemed essential for building optically addressed and cavity-QED based quantum computing nodes~\cite{Kimble2008}. This has particularly spurred research in III-V semiconductor quantum dots~\cite{Lodahl2015,Somaschi2016}, color centers in diamond, silicon carbide and 2D materials~\cite{Doherty,Castelletto,Aharonovich}, and organic molecules at low temperature~\cite{Toninelli}.  On the other hand, classical applications of ensembles of emitters for displays, lighting, lasers and as microscopy-tags drive the continuous development of new types of emitters, such as II-VI self-assembled quantum dots 20 years ago~\cite{Murray1993,Talapin,Shirasaki}, and inorganic perovskite quantum dots just recently~\cite{Protesescu2015,Swarnkar2015,Park2015,LiHuang2018,Gibson2018,Seth2016,Yuan2018,HouKovalenko2020}. For all these systems, understanding the photophysics on the single emitter level is instrumental, whether the intended use is at the single or ensemble level.
A common challenge for almost all types of emitters is that they exhibit intermittency, also known as blinking~\cite{Frantsuzov2008,ToddKrauss2010}. Under constant pumping emitters switch, seemingly at random, between brighter and dimmer states, often corresponding with higher and low quantum yield (QY), and different fluorescence decay rates. Frequently the switching behavior also shows peculiar, power-law distributed, random distributions of durations of events. Determining the mechanism through which emitters blink, i.e., the origin of the involved states, the power-law distribution of residence times, and the cause of switching, have been the topic of a large number of studies particularly for II-VI quantum dots as recently reviewed by Efros and Nesbitt.~\cite{Efros2016}.  Recent studies on inorganic perovskite quantum dots uncover intermittency behavior that does not fit common models for intermittency in their II-VI counterparts~\cite{Swarnkar2015,Park2015,LiHuang2018,Gibson2018,Seth2016,Yuan2018,HouKovalenko2020}.

\begin{figure*}
  \includegraphics[width=\linewidth]{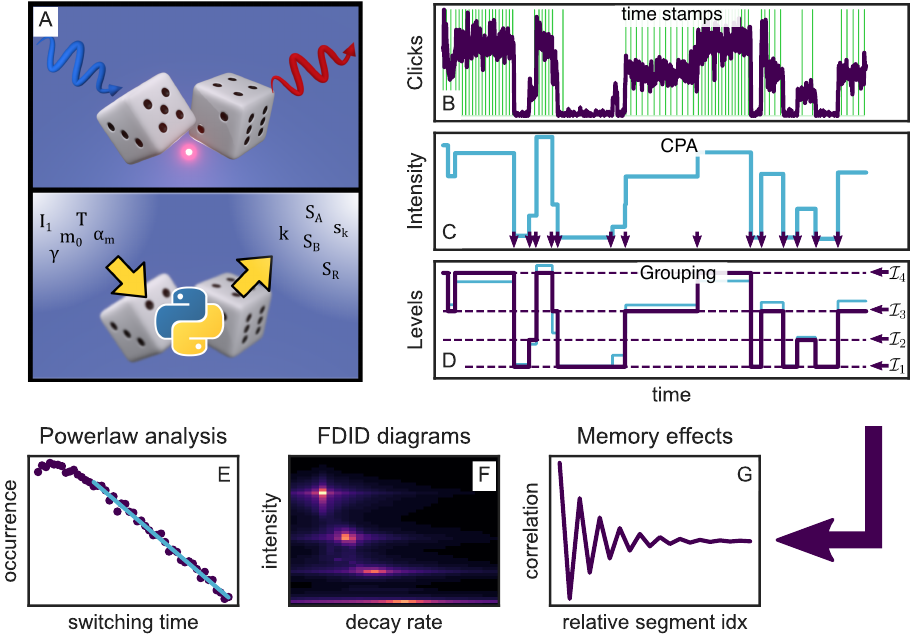}
\caption{A schematic overview of the working of the toolbox. (A) An illustration of the two methods available to obtain TCSPC data, either by photolumiescence TCSPC measurements of a single emitter (top panel) or through simulation of a single emitter (bottom panel). The latter is provided in the toolbox. Both will result in a stream of timestamps that can then be further analyzed by the toolkit. The simulation part of the toolbox simulates dots of $m_0$ levels, with associated count rates and fluorescence decay rates $I_m$ and $\gamma_m$,  with the dot visiting levels in random order, and with residence times for each segment chosen drawn according to specified  power law   exponents $\alpha_m$.  The output simulated data consists of time stamped photon arrivals over a total time span $T$, where for each photon $k=1\ldots $, a time stamp $s_k$ is recorded. These time stamps are randomly distributed over  two detector channels $S_A, S_B$. The delays of each of the photon time stamps  relative to the time stamp in the third channel $S_R$, representing the periodic pump laser pulse train,  is chosen in accordance with the set emitter decay rate. (B) Starting out with this stream of photon events, using CPA (C), the changepoints are found, and with these, the instantaneous intensities. (D) Subsequently, these events are grouped in order to find the most likely underlying intensity levels between the behavior. 
Following this, a number of analyses can be done, such as 
(E) ascertaining whether the time between switching events is  powerlaw distributed, 
(F) the visible separation of states in FDID/FLID  (fluorescent decay rate intensity diagrams, resp. fluorescent lifetime intensity diagrams), and 
(G) the presence of memory in the switching behavior.
Here we show an example of a simulated, 4 level emitter, with simulation parameters chosen for  clarity of illustration.}
  \label{fig:generatedata}
\end{figure*}

In order to quantify intermittent behavior, the simplest and most commonly employed method is to subdivide a measurement stream of individual photon-arrival times into short bins of a few ms, to calculate the intensity (in counts/second) of each bin. Every bin can then be assigned to a state (on, off or grey) according to its brightness so that on/off times as well as intensity levels can be defined and analyzed~\cite{Frantsuzov2008}. For pulsed laser excitation, also quasi-instantaneous fluorescence decay rates can be obtained~\cite{Galland2011,Rabouw2013}. However, it is well known that this method of binning time streams and histogramming binned intensities causes detrimental artefacts~\cite{Watkins2005,Hoogenboom2006,Frantsuzov2008,Ensign2009,Ensign2010,Crouch2010,Houel2015,Bae2016}. Retrieved parameters of the quantum dot behavior often exhibit a dependency on the choice of the bin width, which affect estimates of switching time distributions and power laws,and also the objective assignment of intensities to intrinsic levels. Narrower bin widths in principle allow better resolution, but run into shot noise limits, while conversely choosing larger bins suppresses noise, but will render the analysis blind to fast events. To overcome these issues, Watkins and Yang \cite{Watkins2005} proposed changepoint analysis (CPA) as a Bayesian statistics approach for the unbiased determination of switching times that is optimal, i.e., gives the best performance given the constraints of shot noise in the data. \textcolor{black}{Changepoint analysis and clustering is one example of Bayesian inference methods to determine the transitions and underlying levels in single photon trajectories. In the domain of high-throughput single-molecule analysis\cite{Hill} many methods to process single photon trajectories have appeared, that one can classify as supervised learning methods with a priori model assumptions on one hand, and unsupervised approaches on the other hand. Prominent  are socalled hidden Markov Model (HMM) methods \cite{Hadzic} that  view photon data streams as experimentally measured output of transitions between hidden transition states. Bayesian inference can then estimate parameters such as transition probabilities if one a priori  postulates the number of levels and the allowed transitions. As this underlying model is a priori often not known, one can apply HMM with different possible models, and rank them according to probabilistic criteria, such as the Bayesian Information Criterion. Also, the requirement for a priori known models is relaxed in socalled aggregated Markov models \cite{schmid2016}, and non-Markov memory kernel models \cite{Presse2014}.   Juxtaposed to such supervised analysis methods are unsupervised approaches. Such methods apply changepoint analysis to partition data into time segments between jumps, and subsequent clustering of intensity levels.  The CPA method pioneered by Watkins and Yang is essentially such a combination of changepoint detection and hierarchical agglomerative clustering, using  the Bayesian Information Criterion to determine the best clustering of measured intensities in distinct levels (states) with as sole assumption that intensity in each segment of time wherein the emitter is in a given level, the counts are Poisson distributed \cite{Watkins2005,Ensign2010}. A main drawback is that particularly the clustering is slow. We refer to Ref.\cite{disc2020} for recent developments in machine learning to mitigate this problem. A main advantage of CPA  is that no underlying model is required, and that the data is segmented and clustered to the level that the data allows, given that the data is Poisson distributed in intensity, and given a required confidence level stipulated by the user. Variations on CP for other types of noise, such as Gaussian noise, have also appeared\cite{Shuang2014,LiYang2019}. Despite the well-documented superior performance over binning of photon counting data, in the domain of single photon counting data from quantum dots only very few groups have adopted these methods~\cite{Watkins2005,Ensign2009,Ensign2010,EnsignThesis2010,Schmidt2012,Bae2016,Gibson2018}.}

In this paper, we provide, benchmark, and document a Python toolbox for changepoint analysis, state clustering, and analysis of fluorescence-intensity-decay rate correlations that is posted on GitHub\cite{github}. \textcolor{black}{A main motivation lies in the emergence of new quantum emitter systems with complex photophysics. While II-VI quantum dots for which CPA was orginally developed, are generally understood to switch between just two or three states, the problem of accurate analysis of intermittency is gaining in prominence with the advent of novel emitters, such as perovskite quantum dots which appear to switch not between just two, but instead a multitude of states~\cite{Swarnkar2015,Park2015,LiHuang2018,Gibson2018,Seth2016,Yuan2018,HouKovalenko2020}. There is hence a large need for  a toolbox that  provides unbiased,  model-free analysis of photon counting data, for which reason we provide a CPA implementation and benchmark it for complex multilevel emitters.} Our toolbox is both applicable to real quantum dot data, and valuable as a testbed for both testing models and analysis techniques on synthetic, i.e., numerically generated data.  Indeed, the code toolbox we supply includes  code to generate numerically random photon time arrival data streams for `synthetic' quantum dots that jump between an arbitrary set of intensity levels and decay rates, with jump time statistics and photon budgets that can be set by the user. As results we provide benchmarks on the performance of CPA for detecting change points in function of the number of intensity levels and total photon budget, and we explore the limits to the number of distinct states that the clustering analysis can reliably separate. Moreover, the toolbox allows to us test the accuracy of jump time statistics, such as power law statistics, for such multilevel dots. Finally, the test suite also allows to benchmark the accuracy of fluorescence decay model fitting with maximum likelihood estimation, and we discuss the construction of fluorescence-intensity-decay rate correlations from CPA-partitioned data. \textcolor{black}{The applicability of the toolbox to experimental data is illustrated by supplied data sets, which correspond to a related paper in this journal \cite{Palstra2021expt}.  We note that some other approaches  such as HMM methods may be more suited for processes where more knowledge on the underlying physical processes is available. In contrast our toolbox is ideal for cases in which one wants to make no a priori assumption on the physical mechanism behind intermittency. Furthermore, the CPA method in the toolbox operates on the finest level of information available in photon counting, i.e.. the distribution of individual photon arrival times, as opposed to methods that are optimized to work on camera frame data \cite{Hill}.}  This paper is structured as follows. In the Methods section we summarize the Bayesian statistics tools we implemented to analyze all aspects of our data. Next, we benchmark the performance of changepoint analysis to pinpoint intensity jumps, and of level clustering to identify the number of levels between which a dot switches on the basis of Monte Carlo simulations. Next we present considerations on the dependence of on-off time distributions, decay rate fit and so-called `fluorescence decay-rate versus intensity diagrams'(FDIDs) on count rates. 

%%%%%%%%%%%%%%%%%%%%%%%%%%%%%%%%%%%%%%%%%%%%%%%%%%%%%%%%%%%%%%%%%%%%%%%%%%%%%%%%%%

\section{Methods}
In this section we present all the methods implemented in our Python toolbox, as well as the methods for benchmarking them. Benchmark results are presented in the results section. We refer to the supporting information for a manual to the code and the code itself.

\subsection{Changepoint analysis}
First, we summarize changepoint analysis (or CPA), a Bayesian statistics method for the unbiased determination of jumps or `changepoints' in time traces of discrete events \cite{Watkins2005,Zhang2006,Rubinsztein2009,Ensign2009,Ensign2010,Cordones2011,Schmidt2012,Schmidt2014,Bae2016,Gibson2018,Rabouw2019,LiYang2019}. 
Bayesian statistics is a paradigm that reverses the usual standpoint of probability theory. 
Usual probability theory views a data set as a \textit{random draw} from a probability distribution, given a hypothesis on the parameters of the underlying physical process. In this framework, one can calculate the likelihood of drawing the specific measured data set. Bayesian statistics, on the other hand, compares the likelihood of distinct hypotheses, \textit{given} a measured data set and assumptions on the underlying measurement noise. 

We consider time-tagged single photon counting data consisting of an ordered list of measured photon arrival times $s_k$, collected over a measurement time $T$. For a single emitter with no memory that emits at a count rate of $N$ photons in a time $T$, the waiting times - i.e., the times between photon arrivals - are exponentially distributed with waiting time $\tau_w= T/N$. 
In order to determine whether there is a changepoint in some segment $q$, CPA compares the likelihood of two distinct hypotheses, (1) there is a jump in emission intensity (i.e. the average waiting time $\tau_w$ jumping from some value to another) against (2) there is the same intensity throughout the measurement interval. 
When testing for a jump at photon detection event $k$ at time $s_k$ in this trajectory $q$ with time duration $T_q$ containing $N_q$ photon events, this leads to a log-likelihood ratio, or `Bayes factor'~\cite{Watkins2005,Ensign2009,Ensign2010}
$$
{\cal L}_k = 2 k \ln\frac{k}{V_k} + 2(N_q-k)\ln\frac{N_q-k}{1-V_k} - 2N_q\ln N_q,
$$
where $V_k=s_k/T_q$. Derivation of this log-likelihood ratio involves several steps. First, it incorporates the assumption that in between jumps, the waiting time between photons is exponentially distributed, on basis of which one can assess the likelihood of measuring the given data set for a given hypothesis on the exponential waiting time $\tau_w$. Second, it uses maximally non-informative priors for ${\cal L}_k$ to compare the hypothesis of presence versus absence of a changepoint without further restrictive assumptions on the involved intensity levels. 

It should be noted that there are other ways to arrive at the same log-likelihood ratio test. 
One alternative starting point is a binary time series in which there is an underlying uniform and small probability distribution of photon detection per bin (e.g., imagining the time axis binned in by the timing card resolution (of order 0.1 ns for typical hardware)) \cite{EnsignThesis2010}. Such a uniform distribution would emerge as a direct consequence of exponential waiting time distributions. In this case one should start from a binomial distribution and ultimately arrives at the same formula after application of Stirling's formula.
Another starting point is CPA applied to binned data with wider bins with multiple counts, i.e. to series of Poisson distributed intensities instead of discrete events~\cite{Watkins2005,Ensign2010}. However, the binning would introduce an undesirable time scale through the chosen bin width. 
Of these three methods, working with photon arrival times is the most data efficient approach and introduces no artificial partitioning whatsoever. We refer to Ref.~\cite{Ensign2010} for a derivation of the log-likelihood ratio in all these three scenarios, which includes a precise description of the use of maximally non-informative priors.

Following Watkins and Yang~\cite{Watkins2005} and Ensign~\cite{Ensign2010}, the most likely location of a changepoint, if any, is at the $k$ that maximizes the Bayes factor ${\cal L}_k$. The hypothesis that this most likely changepoint is indeed a real event is accepted if ${\cal L}_k$ exceeds a critical threshold value for ${\cal L}_k$ or `skepticism'. This value is chosen to balance false positives against missed events. A full data set is partitioned recursively, i.e., by recursively checking if data sets between two accepted changepoints themselves contain further changepoints. This results in a division of the data set into segments, each of which starts and ends at an accepted changepoint, and with the level of skepticism as stop criterion for the recursion. The resulting segmentation provides the most likely description of data as consisting of segments within which the intensity is constant, given the value chosen for the degree of `skepticism', and given the amount of data collected. Since the algorithm works with the list of individual photon arrival times this segmentation entails no arbitrary partitioning. \textcolor{black}{An accepted rule of thumb is that if the Bayes factor ${\cal L}_k$ exceeds a `skepticism' value of just between 1-3,  the evidence for a changepoint is highly ambiguous, while values in the range 7-10 are deemed strong evidence The toolbox is supplied with a default value of skepticism of 8, set following the analysis of~\cite{Watkins2005}  and ~\cite{Ensign2010}.  The reader is warned that for a given photophysics scenario (intensity levels, segment duration statistics) it is advisable to set the level of `skepticism' on basis of simulations, in order to optimize the trade off between missing changepoints altogether (false negatives) and precision (avoiding false positives). Our result section provides an example of such an optimization.}

\subsection{Clustering}
\label{subsec:method_clustering}
Changepoint analysis splits the data in segments separated by jumps  (a list of $Q$ jumps delineate $Q-1$ segments). One can now ask what the statistical properties are of the segmentation, i.e., what the statistics is of the length of segments, the intensity levels most likely corresponding to the segments, and the fluorescence decay times associated with the segments. For instance, it is a nontrivial question how many distinct constant intensity levels, or states, $m_r$ actually underlie the $N-1$ found segments, with intensities $I_1 \ldots I_{Q-1}$. To answer this question Watkins and Yang~\cite{Watkins2005} proposed a clustering approach. The recent work of Li and Yang~\cite{LiYang2019} provides a detailed explanation of the reasoning involved, though quoting results for Gaussian instead of Poissonian distributed data. 
The idea is that with the $Q-1$ found segments, each with their associated recorded intensities $I_q$ one can use expectation maximimization  to calculate, for a hypothesized and fixed number of levels $\nG$ what the most likely underlying intensity levels $\mathcal{I}_{m}$ are (with $m \in 1 \ldots \nG$), and how probable it is that each segment is ascribed to a given level (probability $p_{mq}$).  Subsequently Bayesian inference is used to establish what the most likely number of levels (i.e. states) $m_r$ and associated intensities $\mathcal{I}_{m}$, with $m \in \{ 1 \ldots m_r \}$ is that describes the data.
 
Following Ref.~\cite{Watkins2005,LiYang2019}, the expectation minimization in our toolbox is implemented as an interative algorithm started by a first guess of the segmentation. This guess is obtained by a hierarchical clustering of $Q-1$ segments in $m=1,2,\ldots Q-1$ levels that proceeds recursively. In each step it identifies the two segments in the list with the most similar intensity levels as belonging to the same level. This provides an initial clustering of the measured data in any number $m=1,2,\ldots Q-1$ of levels. For the expectation maximization the idea is to simultaneously and iteratively optimize the probability $p_{mq}$ for segment $q$ to belong to the $m$th level, as well as an estimate of the intensities of these levels $\mathcal{I}_{m}$. In each iteration the intensities of all levels are estimated from the level assignment from $p_{m,q}$. Following this, the probability distribution $p_{mq}$ is updated to redistribute the segments over the levels.
In this calculation it is important to understand the type of noise statistics the data obeys. In the case of single-photon measurements, and for the purpose of this discussion, the intensities are Poisson distributed.  The iteration is repeated until $p_{mq}$ converges (practically also capped by a maximum number of iterations). The final outcome is a most likely assignment of the measured segments into $\nG$ levels. Next for each value of $nG$ one asseses the  `Bayesian information criterion' (BIC). This criterion is a measure for how good the description of the segmented intensity trace is with $\nG$ intensity levels given the assumption of Poisson counting statistics for each fixed intensity level. Beyond a mere `goodness of fit' metric that would simply improve with improved number of parameters available to describe the data, this metric is penalized for the number of parameters to avoid overfitting. For Poisson distributed data the criterion is derived in\cite{Watkins2005} as  
$$
\textrm{BIC} = 2{\cal L}_{EM} - (2\nG-1)\ln Q - Q\ln N
$$ 
where $Q$ again is the number of change points detected, $\nG$ is the
number of available levels. The term  ${\cal L}_{EM}$ is the log-likelihood function optimized in the expectation maximization step, i.e ${\cal L}_{EM} = \sum_q\sum_{m=1}^{\nG} p_{mq} \text{ln}[p_m{\cal P}(I_q;{\cal I}_m)]$ with ${\cal P}_(x;\lambda)$ the Poisson probability function at mean $\lambda$, $p_m$ the probability of drawing level $m$. The second term in the BIC s the term penalizing the BIC for overfitting. The accepted best description of an emitter in $\nG$-levels is taken to be at the value of $\nG$ where the BIC peaks.

\subsection{Intensity cross/autocorrelation and maximum likelihood lifetime fitting}
Many single photon counting experiments are set up with pulsed laser excitation for fluorescence decay rate measurements, and with multiple detectors to collect intensity autocorrelations (e.g., to verify antibunching in $g^{(2)}(\tau)$ for time intervals $\tau$ comparable to the fluorescence decay rate, and shorter than the commonly longer detector dead time). In a typical absolute-time tagging set up, this results in multiple data streams $S_\text{A}, \, S_\text{B}, \, S_\text{R}$ of time stamps corresponding to the detection events on each detector, and the concomitant laser pulses that created them, respectively. Our Python toolbox contains an implementation of the correlation algorithm of Wahl et al.\cite{Wahl2003} that operates on timestamp series, and returns, for any combination of channels $S_1, S_2$ ($1,2 \in \{ \text{A,B,R} \}$), the cross correlation $C(\tau)\Delta\tau$, i.e., the number of events in the time series $S_1$ and the time series $S_2$ that coincided when shifted over $\tau$, within a precision $\Delta\tau$.

Cross-correlating detected photons and laser arrival times, taking $\Delta\tau$ to be the binning precision of the counting electronics and the range of $\tau$ equal to the laser pulse repetition rate, returns a histogram of the delay times between photon detection events and laser pulses. 
To obtain $g^{(2)}(\tau)$ to investigate antibunching, streams of photon events from two detectors in a Hanbury-Brown Twiss set up are cross-correlated. $\Delta\tau$ is taken to be the binning precision of the counting electronics and the sampled range of $\tau$ as an interval is taken symmetrically around $\tau=0$ and several times the laser pulse interval.
 Finally auto- or cross correlating detector streams over $\tau$-ranges from nanoseconds to seconds, coarsening both $\tau$ and $\Delta \tau$ to obtain equidistant sampling on a logarithmic time axis, results in long time intensity autocorrelations of use in intermittency analysis~\cite{Houel2015}. Our toolbox also provides this logarithmic time-step coarsening version of the correlation algorithm of Wahl et al.~\cite{Wahl2003}

Of particular interest for intermittent single emitters is the analysis of fluorescence decay rates in short segments of data as identified by changepoint analysis, that may be so short as to contain only 20 to 1000 photons. For each of the photon detection events in a single CPA segment, cross-correlation with the laser pulse train yields a histogram of the $N_q$ photons in segment $q$. 
In each of the bins (with width $\Delta\tau$) the photon counts are expected to be Poisson distributed. Therefore the optimum fit procedure to extract decay rates employs the Maximum Likelihood Estimate procedure for Poisson distributed data, as described by Bajzer et al.~\cite{Bajzer1991}. 
In brief, for a decay trace sampled at time points $\tau_i$ relative to the laser excitation, with counts per bin $D(\tau_i)$, the merit function reads
\begin{equation}
\label{Eq_probabSum}
  M=
  -\sum_{\text{all data bins $i$}} \left\{ D(\tau_i)\,\text{log}\left[F_{A}(\tau_i)\right] - F_{A}(\tau_i) \right\}.
\end{equation}
Assuming a chosen fit function $F_A(t)$ the parameter set $A$ that minimizes this merit function provides the parameter values that most likely correspond to the data. The estimated errors in these parameters then follows from the diagonal elements of the inverse of the Hessian of $M$ relative to the parameters $A$. Importantly, the fact that the Poisson distribution is tied to absolute numbers of counts, implies that this approach requires that the data is \textit{neither} scaled nor background subtracted. Instead, the background should be part of the fit function either as a free parameter or a known constant. Furthermore, it should be noted that time bins with zero counts are as informative to the fit as non-empty ones, and should not be left out.

\subsection{Generating synthetic quantum dot data}
To benchmark the CPA and clustering method and to test its limits, our toolbox provides an example routine to generate artificial data mimicking quantum dot intermittency. To obtain mimicked quantum dot data, we first choose a number $m_0$ of intensity levels $\mathcal{I}_{0,m}$ between which we assume the dot to switch. Next we generate switching times for each of the states. In this work, we choose all switching times from a power law distribution. For benchmark purposes we will present results with power law exponent $\alpha = 1.5$, though any exponent can be set in the code. On the assumption that intensity levels appear in a random  and uncorrelated order, this segments the time axis in a list of switching events $T_{0,j}$, $j=1,2,\ldots m_0 $, where for each segment we randomly assign one of the nominal intensities $\mathcal{I}_{0,m}$. Next, to mimic a pulsed excitation experiment, we imagine each of these segments to be subdivided in intervals of length $\tau_L$ equivalent to a laser repetition rate ($\tau_L=100$ ns in the examples in this work). We assign each of these intervals to be populated with one photon at probability $p_m=\mathcal{I}_{0,m}\tau_L$. This ensures that the number of photons $N_q$ in every segment is drawn from a Poisson distribution at mean $T_{0,q} \mathcal{I}_{0,m}$. By removal of all empty bins, the binary list is translated into a list $S=(t_1,t_2,t_3,\ldots)$ of photon arrival time stamps at resolution $\tau_L$ to which one can directly apply CPA to attempt a retrieval of switching times, and apply clustering to retrieve the number of states.

To also enable fluorescence decay rate analysis, we further refine the photon arrival time list. Recalling that we have generated switching events $T_{0,q}$ between intensity states $\mathcal{I}_{0,m}$, we now also assume fluorescence decay rates $\gamma_{0,m}$. As each segment $q$, was already chosen to correspond to some level $m_q$, we now impose decay rate $\gamma_{0,m_q}$ on the photon arrival times. To do so, for each of the photon events $k=1\ldots N_q$ already generated at resolution $\tau_L$ we now randomly draw a delay time $\Delta_k$ relative to its exciting laser from an exponential distribution characterized by rate $\gamma_{0,m_q}$. To mimic the behavior of typical TPSPC counting equipment, the delay time is discretized at a finite time resolution $\Delta\tau$ (in this work chosen as 165 ps to match the hardware in the provided example experimental data measured in our lab (Becker \& Hickl DPC-230), though of course in the toolbox the value can be set to match that of any TCSPC card vendor). 
Testing of fluorescence decay trace fitting can operate directly on the generated list of delay times, or alternatively one can synthesize a TCSPC experiment by re-assigning $S$ to represent laser-pulse arrival times $S_\text{R}$, and defining photon arrival times as the events in $S_A$, $S_B$ each shifted by its delay time, i.e. $S_\text{X}=(t_1+\Delta_1, \, t_1+\Delta_2,\ldots)$,with $\text{X}\in\text{A,B}$. Cross correlation of $S_\text{R}$ and $S_\text{X}$ returns the delay time list. We note that although our work does not focus on antibunching, our quantum dot simulation routine provides data distributed over two detector channels, where emission events antibunch, while an uncorrelated background noise level of the detectors can also be set. 

\subsection{Practical implementation}
We have implemented the toolbox ingredients in Python 3.8. As timestamp data can be substantial in size, we use the `parquet' binary format to store timestamps as 64-bit integers.  Processing and plotting the data is dependent on  Pythons' standard libraries numpy, matplotlib, while we use Numba, a just-in-time compiler, to accelerate the time-stamp correlation algorithms. An example script to generate synthetic data, and to run the entire workflow on simulated data is provided.  We refer to the supporting information for a guide to the practical implementation and use of our toolbox.   The toolbox comes  also with  example experimental data on single CsPbBr$_3$ quantum dots from a recent experiment~\cite{Palstra2021expt}{\bfseries under review at the same journal for intended back to back publication}.    Table \ref{table} list scaling and performance metrics for the algorithms contained in the toolbox.

%%%%%%%%%%%%%%%%%%%%%%%%%%%%%%%%%%%%%%%%%%%%%%%%%%%%%%%%%%%%%%%%%%%%%%%%%%%%%%%%%%

\begin{table}
\begin{tabular}{|l|c|c|}
\hline
\textbf{Algorithm}        & \textbf{Scaling}   & \textbf{Timing}                                                                                                                     \\
\hline
\hline
CPA                       &      ${\cal O}(N\sqrt{N})  $            & Ca. 10 s for $N=10^6 $                                                                                                                                                     \\
\hline
Grouping                  & \multicolumn{2}{l|}{Initial clustering dominates over iterative algorithm}                                                                                                                         \\
Initialization\footnotemark[1]                   &      $ {\cal O}({Q^2})   $         & 0.5 s for $Q=3\cdot 10^2 $                                                                            \\
Iterative optimization\footnotemark[2]                 & Overhead dominated & 0.75 ms per iteration ($Q< 500$)                                                              \\
\hline
$g^{(2)}$\footnotemark[3]&          ${\cal O}(N n_\text{plot points})$          & 8 ms per plot point at $N=10^6$  \\
\hline
Long-time autocorrelation\footnotemark[4] &   ${\cal O}(N n_\text{cascade})$                 & Ca. 2s for full plot                                                              \\ 
\hline
\end{tabular}
\flushleft
\footnotesize
\footnotemark[1] For large Q this is accelerated by first clustering subsets,  merging, and continuing clustering\\  
\footnotemark[2] Typically 5 to 10 iterations are required when  $n_G\approx m$.  \\
\footnotemark[3] Numba JIT acceleration assuming int64  provides over 2 orders of magnitude accelation. A typical $g^{(2)}$ plot has ca.  $n_\text{plot points})=2000$, and hence requires 10 to 20 seconds to evaluate.\\
\footnotemark[4] Essentially repeating $g^{(2)}$ and logarithmic coarsening every $n_\text{cascade}$ points.               \\
\caption{Algorithm scaling and computation time as function of number of photons $N$ and changepoints $Q$. Timings were obtained on a standard desktop [Intel I7 4790 at 3.6 GHz, with 16 Gb of DDR3 RAM), and are obtained on basis of 5 $\times$ 100 photon trajectories (100 independent draws for 5 different trajectory record lengths with from ca. $10^4$ to $3\cdot10^6$ photon events).\label{table}}
\end{table}

\section{Results}
The remainder of this work is devoted to presenting benchmarks of the provided methods. Benchmarks for emitters with `binary'  switching, i.e.,  two  well-separated intensity levels as is typical for  II-VI quantum dots, have already been presented in literature~\cite{Watkins2005,EnsignThesis2010,Bae2016}.  However, emitters under current study, such as perovskite quantum dots appear to have a multitude, or perhaps even a continuum, of intensity levels. Our tests hence focus on determining the performance of CPA and level clustering for many-level single photon emitters. 

\subsection{Precision of identifying individual changepoints}
\label{subsec:method_fakeqdot_grouping}

\begin{figure*}
  \includegraphics[width=\linewidth]{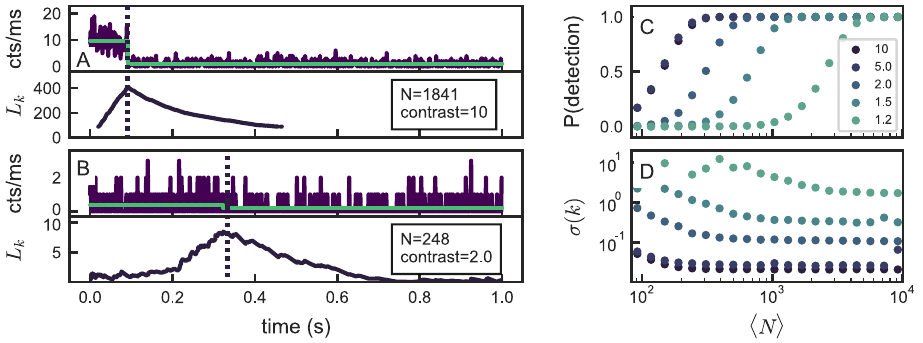}
\caption{Demonstration of changepoint detection applied to a synthesized data set with a single changepoint, with equal photon counts before and after the changepoint. In  (A) and (B) the contrast between intensities is a factor of 10 and 2, respectively, while the total photon budget is approximately 2000 and 300. The bottom panels show the log-likelihood ratio test, which clearly peaks at the changepoint in both cases. The y-axis unit cts/ms stands for counts per millisecond. The robustness of the method is demonstrated in (C) and (D), where we show the likelihood of detecting a changepoint in such a series for different intensity contrasts, and the variance of the found times, respectively, as a function of the total photon numbers. To gather accurate statistics, $10^4$ photon traces were generated for each data point. The data is plotted in a binned fashion (ms bins) for visualization purposes only.}
  \label{fig:sim_likelihoods_detect}
\end{figure*}
Figure~\ref{fig:sim_likelihoods_detect}(A) and (B) show examples of CPA analysis applied to a simulated quantum dot with a single jump in its behavior, from an intensity level $10^4$ to $10^3$~cts/s resp. from $4.5\times10^2$ to $2.25\times10^2$, with $\sim 900$ and $\sim 150$ photons left and right of the changepoint, respectively.
Purely for visualization purposes, the data is plotted in a binned format, as the analysis itself does not make use of any binning.
Alongside the binned intensity trace, we also show the log-likelihood ratio ${\cal L}_k$. In both cases, the log-likelihood ratio clearly peaks at or close to the point where there is a changepoint in the data. Since the Bayes factor is actually a \textit{logarithmic} measure for the comparison of hypotheses, the algorithm indeed identifies the changepoint with high probability and to within just a few photon events, even where the jump is far smaller than the shot noise in the binned representation in the plot, at a relative intensity contrast of just a factor of 2. Generally, the probability with which the algorithm identifies or misses the changepoint is dependent on the total number of photons recorded both before and after the changepoint, and on the contrast in intensities, consistent with the findings of Watkins, and Ensign~\cite{Watkins2005}. 

To identify the limits of CPA~\cite{Ensign2010,Bae2016} we consider the feasibility of identifying changepoints of contrast $I_2/I_1$ as function of the total number of photons in the time record. The results are shown in {Figure~\ref{fig:sim_likelihoods_detect}C} for the likelihood of detecting a changepoint, and {Figure~\ref{fig:sim_likelihoods_detect}D} for the error in identifying the precise event $k$ at which the changepoint that is identified occurred. Here, we only consider the case where there are roughly an equal number of photon events before and after the changepoint. This data is obtained by simulating $10^4$ switching events of the type as shown in {Fig~\ref{fig:sim_likelihoods_detect}(A,B)} for each contrast and mean photon count shown. \textcolor{black}{The range of contrasts is chosen commensurate with reported on-off contrasts for typical quantum dots in literature, which generally fall in the 1.5 to 5-fold contrast range.} At high intensity contrast, exceeding a factor 5, a total photon count as low as 300 is enough for near-unity detection. Moreover, for sufficiently high photon count left and right of the changepoint, even very small changes in intensity have a high likelihood of being accurately detected, even if in binned data representations the jump is not visible within the shot noise. {Figure \ref{fig:sim_likelihoods_detect}(D)} provides a metric for the accuracy to within which changepoints are pinpointed. Changepoint analysis returns the most likely photon event $k$ in which the jump occurred, which in our analysis can be compared to the actual photon event index $k_0$ at which we programmed the Monte Carlo simulation to show a jump.{Figure \ref{fig:sim_likelihoods_detect}d} reports the mean error ($\sqrt{\text{var}(k-k_0}$) as a metric of accuracy. At jump contrasts above a factor $2$, changepoints are identified to within an accuracy of almost one photon event even with just $10^2$ photon counts in the total event record. At very small contrasts, the error in determining the location of a changepoint is generally on the level of one or two photon events, only worsening when there are fewer than 200 counts. This observation highlights the fact that if the photon record has just a few counts in total, the error in estimating the count rate before and after the jump becomes comparable to the magnitude of the jump. 

\subsection{Intermittency and on-off time histograms}
\begin{figure*}
  \includegraphics[width=\linewidth]{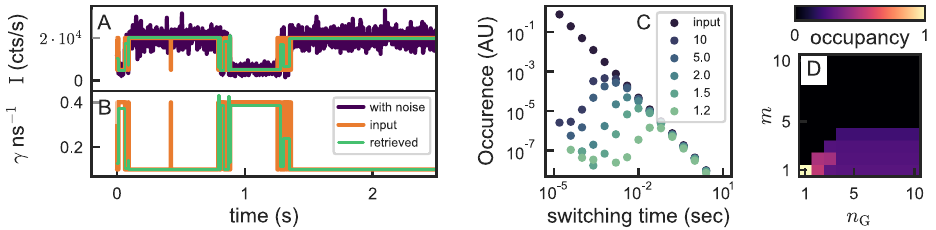}
  \caption{
  (A) Typical time trace of a simulated quantum dot. The intensity duty cycle switches between $0.5\times 10^4$ and $2\times 10^4$ counts/s. It shows an on/off input duty cycle generated with a power-law distribution (orange), the duty cycle with Poissonian noise (purple), and the retrieved duty cycle (green). Overall, the original intensities and lifetimes (B) are retrieved well. 
  (C) A histogram of the number of switching events as a function of their duration. Each data point represents 10.000, 10 s power-law distributed time traces. The input shows the initial power-law distribution, the lighter colors show the number of retrieved changepoints, after applying Poisson noise and CPA, at different contrasts, with $I_1=10^5$. counts/s. We can see that even at low contrast, events with long times between switching are retrieved, but each contrast has a characteristic duration below which changepoints can't be accurately retrieved. This puts a fundamental limit on the information that can be extracted from a given data set. 
  (D) An occupancy diagram illustrating the behavior of the clustering algorithm for a system with $m_0$=4 intensity levels. The color scale indicates the amount of time spent in each state $m_i$ after the assignment of states for a given number of available states $\nG$. We see that when $\nG > m_0$, effectively all segments are distributed across only $\nG \leq m_0$ intensity levels.
  }
  \label{fig:sim_timetrace}
\end{figure*}
As next step in our Monte Carlo benchmarking we turn to time series with many, instead of single, jumps. {Fig.~\ref{fig:sim_timetrace}A} shows a representative example for a simulated intermittent quantum dot with two states, assuming a contrast ratio between states of $2\times10^4$ and $5\times10^3$ $\textrm{s}^{-1}$. We generally observe that the recursive CPA algorithm accurately identifies switching events, barring a number of missed events of very short duration. From the CPA analysis we retrieve the time duration between switching events. {Figure~\ref{fig:sim_timetrace}C} shows a histogram of time durations, plotted as a probability density function obtained from a whole series of Monte Carlo simulated time traces of varying contrast between states (see legend). 
Noteably, if we simulate quantum dots that have switching times that are power-law distributed, the retrieved distribution indeed follows the assumed power-law particularly for long times. At shorter times, the histogram remains significantly below the power law, particular at low intensity ratios between the two assumed states. This indicates that CPA misses fast switching events, and is consistent with the observation from {Figure~\ref{fig:sim_likelihoods_detect}C} that a minimum photon count is required to observe switching events of a given contrast. As a rule of thumb, usual II-VI colloidal quantum dots have a contrast between dark and bright states of around $5$, meaning that of order 200 photons are required to detect a change point with near-certainty. At the assumed count rates ($2\times10^4 \, \textrm{s}^{-1}$ for the bright state) this means one expects CPA to fail for switching times below $10$ ms, where the on-off time histogram indeed shows a distinct roll off. This result suggests one should interpret on-off time histograms from changepoint detection with care: one can generally rely on the long-time tail, but should determine the shortest time scale below which the histogram is meaningless on basis of the intensity levels present in the data.

\textcolor{black}{
\subsection{Error analysis for trajectories with multiple jumps}
The changepoint analysis results in Figure \ref{fig:sim_likelihoods_detect}(C) essentially quantifies the algorithm performance in terms of the fraction of correctly identified change points (true positives) for traces with a single step in intensity, as function of contrast and photon budget.  Actual single emitter photon trajectories have a plethora of steps, where CPA is mainly likely to miss short segments due to the fact that only those changepoints are accepted for which the evidence in the data is compelling, relative to the shot noise in it. Indeed, the short-time roll-off in Figure~\ref{fig:sim_timetrace}C highlights exactly this tendency of CPA to under-report on closely-spaced changepoints (false negative rates high for short segments).  The level of skepticism set as parameter for running CPA sets the overall accuracy of the algorithm, essentially trading off the rates of false positives, and false negatives. When using the toolbox for a particular photophysical scenario, the reader is recommended to study the error rates as function of skepticism. To demonstrate that type of study, here we report on algorithm performance as function of skepticism using the error metrics \emph{accuracy}, \emph{precision} and \emph{recall}. To this end,  we generate synthetic data and match the list of nominal changepoints and retrieved changepoints to determine the rate TP of true positives, the rate FP (false positives) of detected transitions for which no transition was actually present, and the rate FN of false negatives, in which a true transition is not detected by CPA.  The standard  definition for the error metrics reads\cite{Hadzic,disc2020}
\begin{eqnarray}
\text{accuracy}&=&\frac{\text{TP}}{\text{TP+FN+FP}}\\
\text{precision}&=&\frac{\text{TP}}{\text{TP+FP}}\\
\text{recall}&=&\frac{\text{TP}}{\text{TP+FN}}.
\end{eqnarray}
The accuracy benchmarks overall performance, whereas precision measures the false positive error rate, and recall quantifies the false negative rate. Since firstly  changepoint detection is not accurate to the level of a single photon arrival time, and secondly the set of stored nominal switching times in our toolbox may fall in between synthsized photon events, such a  comparison requires a tolerance range to be meaningful.    Figure \ref{fig:erroranalysis} presents the algorithm performance as function of the level of skepticism (vertical axis), and as function of the tolerance range within which change points are accepted as true positives, measured in milliseconds. The results are for a more challenging case than a two-level dot, namely a 4 level system with mean count rate $5\cdot10^{4}$ counts per second, and 4 equidistant intensity levels (2, 4, 6 and 8$\cdot10^4$ counts per second),  and power law distributed segment residence times (exponent 1.5, with shortest residence time of 10 ms in a segment). Presented results are obtained from 200 photon trajectories with on average $5\cdot 10^5$ photons and $10^2$ changepoints each.
 If the time axis for the tolerance is chosen as short as the inverse mean count rate,  the apparent algorithm precision is low, indicating that changepoints are generally found close to, but not quite at, the moment where the switching event occurs.  At tolerances of 2 to 5 ms (containing of order 50-250 photons typically at the given rate, and for the various assumed intensity levels) the error rate saturates at above 90\%. The \emph{accuracy} for this example peaks at a skepticism of ca. 7.0 ( Figure \ref{fig:erroranalysis}A). At higher levels of skepticism,  the \emph{precision} increases, i.e., the number of false positives reduces further  (Figure \ref{fig:erroranalysis}B). However, this is at the expense of recall, i.e., the number of missed changed points. The false negatives rate decreases only with skepticism lowered to below 10 (Figure \ref{fig:erroranalysis}C).
 \begin{figure*}
 \includegraphics[width=\linewidth]{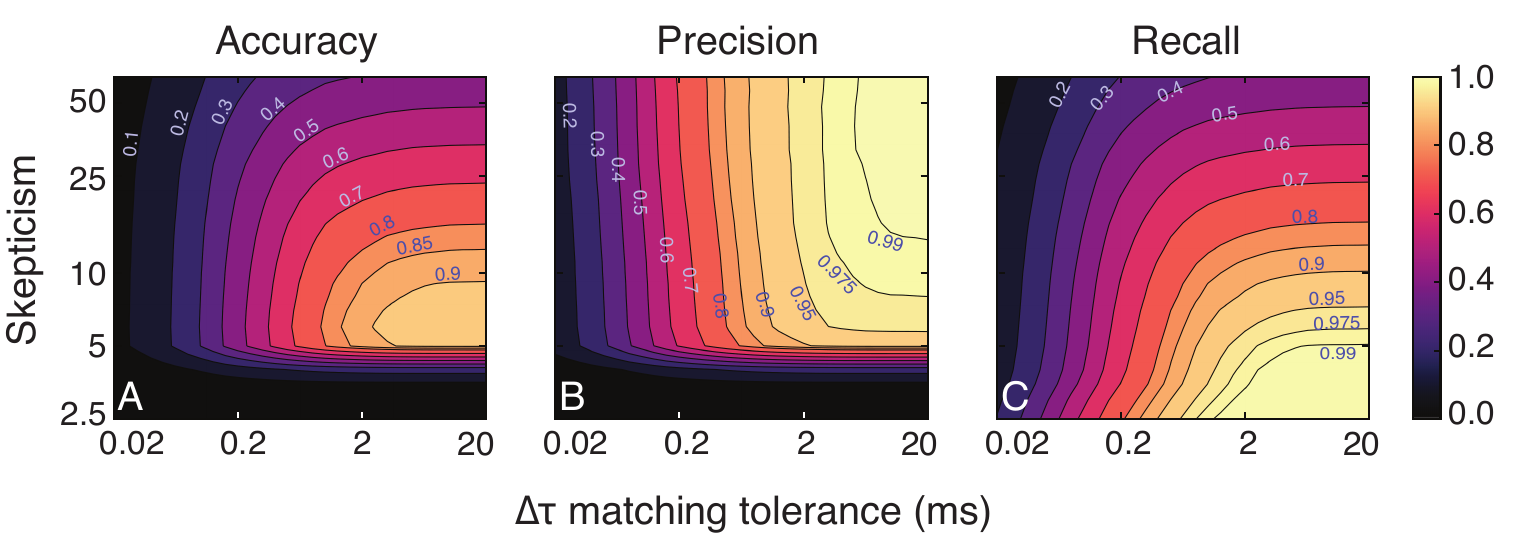}
 \caption{Accuracy (A), precision (B) and recall (C) for changepoint analysis as function of tolerance in milliseconds for which nominal and detected changepoints are matched as equal, and as function of the level of "skepticism" which the Bayes factor needs to exceed for a changepoint to be accepted.\label{fig:erroranalysis}}
 \end{figure*} 
}

\subsection{Performance of level clustering}
\begin{figure*}
  \includegraphics[width=\linewidth]{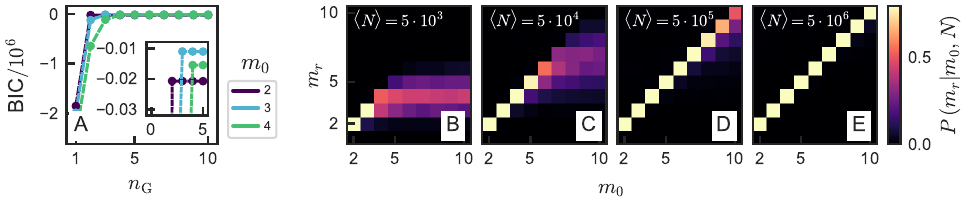}
  \caption{
  (A) Examples of the BIC for simulated quantum dots, with $m_0=2,3,4$. The most likely number of levels is indicated by a small peak in $L_k$. In the inset, we show that the distributions indeed peak at their respective $m_0$. It should be noted that the BIC shows a very sharp rise, and then from $m_0$ onwards appears almost flat. There is in fact a shallow downward slope. Here one should keep in mind that the BIC is a logarithmic metric. On a linear scale the maximum is significant.  (B-E) Likelihood $P$ of finding $m_r$ levels for a simulated system, given $m_0$ initial intensity levels and $\langle N \rangle$ the mean number of photon counts. We see that the photon budget plays a defining role in the total number of states that one can reliably resolve. At low photon budgets, the number of levels is systematically underestimated, whereas at high photon budgets, $P(m_i=m_0)$ remains high even at high $m_0$. We see that $m_r$ is never overestimated.}
  \label{fig:sim_grouping_likelihood}
\end{figure*}

Next we consider the performance of the grouping algorithm applied to the segmentation of simulated time traces. For reference, {Figure~\ref{fig:sim_grouping_likelihood}A} shows the Bayesian Information Criterion versus $\nG$ for the example of  simulated dots with $m_0=2, 3, 4$ intensity levels. Generally, the BIC rapidly rises as $\nG$ approaches the actual number of levels with which the data was simulated, and gently decreases once $\nG$ exceeds the actual number of levels in the data $m_0$. The fast rise indicates that within the assumption of Poisson distributed intensities, the data can not at all be described by fewer than $m_0$ levels. The slow decrease is due to the penalization of the BIC by the number of fit parameters. Since the BIC criterion actually relates to the \textit{logarithm} of the probability with which $\nG$ states are the appropriate description of the data, even an apparently gentle maximum in BIC actually coincides with an accurate, unique determination of $m_0$. 

To gauge the accuracy of the retrieval of the number of states for multi-state quantum dots, we simulated quantum dot data with power-law distributed ($\alpha=1.5$) switching events, assuming switching between from $\tilde{m_0}=2$ to 10 equally likely levels, where we assumed intensity levels to be assigned to segments randomly, and where we assumed levels for simplicity evenly spaced from dark to bright. Lastly, all segments were reassigned an intensity according to Poisson statistics. In other words, we added shot noise.

For many random realizations with different $m_0$ and $\langle N \rangle$ we determine the most likely number of states $m_r$ ($\text{BIC}(m_r) = \text{max}(\text{BIC})$) according to the clustering algorithm, and construct histograms of outcomes. The total photon budget is set by the product of assumed record length and the mean count rates of the different levels. The outcome of these calculations are shown in {Figure~\ref{fig:sim_grouping_likelihood}(B-E)}, where each panel corresponds to a different photon budget. A plot with only diagonal entries signifies that the number of levels retrieved by the clustering algorithm always correspond to the number of levels assumed, so $m_r=m_0$. 
At high photon budgets {(Figures~\ref{fig:sim_grouping_likelihood}D,~E)}, the retrieval of the number of states is indeed robust, even for simulated dots that switch between as many as 10 intensity levels. At low total photon budgets {(Figures~\ref{fig:sim_grouping_likelihood}B,~C)}, we see that $m_r$ is often underestimated. This signifies that there is high uncertainty due to shot noise in the assigned intensity levels, so that levels can not be discriminated within the photon budget.  \textcolor{black}{It is remarkable that at photon budgets of $10^6$ photons, as many  as 10 intensity levels can be robustly discerned even though the smallest contrast between levels is as small as 10\% in intensity in view of Fig. \ref{fig:sim_likelihoods_detect}, where it is evident that detecting changepoints for small intensity jumps is difficult. Here, however, one should realize that in contrast to Fig. \ref{fig:sim_likelihoods_detect}C, where single small-contrast changepoints are studied,  here  many levels are visited in a random order. Thus the correct detection of small level differences is not reliant on the detection of small changepoint contrasts, but on having sufficient photon statistics to resolve the segment count rates of already identified segments.}  
As a secondary metric, additional to the BIC, is in how the clustering algorithm assigns occupancy to the levels. The clustering algorithm assigns to each data segment   the most likely intensity level that it was drawn from.  Occupancy is a metric for how often each of the $\nG$ levels available to the algorithm is actually visited by the measured intensity sequence. 
We find that if one allows the clustering algorithm to use more levels than originally used to synthesize the quantum dot data ($\nG > m_0 $), the additional levels take essentially no occupancy.
We show this in {Figure~\ref{fig:sim_timetrace}D} for an example of a dot assumed to have four intensity levels with a total photon count of $5\times 10^5$. As soon as additional states ($m\geq 5$) are offered to the grouping algorithm, these additional states take no occupancy and do not change the distribution of segments over the states found at the correct $m$. Thus we confirm again that the grouping algorithm does not over-estimate the number of states.

\subsection{Accuracy of decay rates versus photon budget}
\label{subsec:method_decaytraces}
\begin{figure}
  \includegraphics[width=\linewidth]{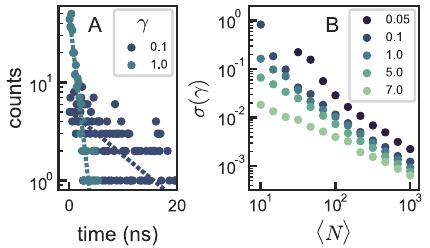}
  \caption{
  (A) Two examples of decay traces ($\gamma=0.1, 1$ ns$^{-1}$) with low photon count ($N=300$) fitted to a single-exponential decay.
  (B) The standard error of the fitted decay rate w.r.t. the input decay rate as a function of the total photon count, for different decay rates. Each data point is the average of $10^3$ simulated decay traces. }
  \label{fig:lifetimes_examples}
\end{figure}

{Figure~\ref{fig:lifetimes_examples}(A)} shows examples of fitted simulated data for slow and fast decays, as examples of the Monte Carlo simulations we have performed to benchmark the accuracy of decay rate fitting in function of photon budget, and decay rate (panel (B)). 
We find that the error in $\gamma$ very roughly follows roughly a power law with an exponent of $0.9-1.1$, with slower decay rates showing higher errors. 
Consistent with Ref.~\cite{Kollner1992} we find by Monte Carlo simulation that one requires approx. 200 (50) counts to obtain an error below 10\% (20\%) in decay rate if one fits mono-exponential decay with free parameters. 
A problem intrinsic to the use of CPA is that fast switching events may be missed, leading to an averaging of short time intervals with others. This leads to decay traces that are in fact not attributable to a single exponential decay.

\subsection{FDID diagrams}
\begin{figure*}
  \includegraphics[width=\linewidth]{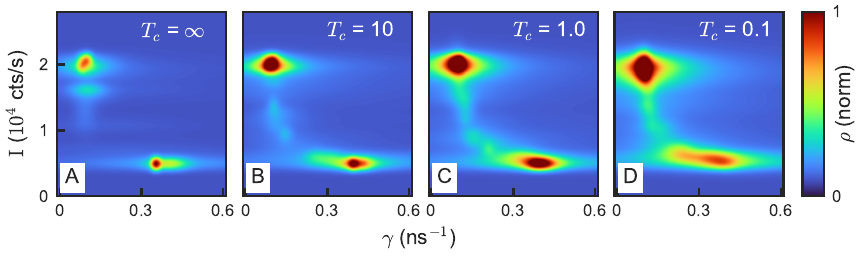}
  \caption{
  Four FDID diagrams of a simulated bimodal quantum dot with ($I,\gamma$) = ($2\times10^4$ s$^{-1}$, 0.1 ns$^{-1}$) and ($0.5\times10^4$ s$^{-1}$, 0.4 ns$^{-1}$). In each diagram, we apply a different cutoff time $T_c$ to the simulated powerlaw at $\infty$, 10 s, 1 s, and 0.1 s. We find that a shorter cutoff causes an increasingly strong smearing effect between the two states.}
  \label{fig:FDIDs}
\end{figure*}

Correlative diagrams  that plot correlations between intensity levels and fluorescence lifetime~\cite{Bae2016, Galland2011, Rabouw2013} form a powerful visualization of quantum dot photophysics.\textcolor{black}{Our toolbox contains code to generate both fluorescence decay rate intensity  diagrams (FDIDs),  and fluorescence lifetime intensity diagrams (FLIDs).  The considerations in this section hold equally for FDID and FLID diagrams, although the provided example is for a FDID analysis}. Essential to the construction of FDID/FLID diagrams is that for each detected photon also the delay time to the laser pulse that generated it is known so that decay rates can be fitted even to short segments of a time trace segmented by CPA. Here we discuss the construction of FDID diagrams derived from CPA-analysis, again illustrated by examining simulated data for a quantum dot that switches between two states of distinct intensity and lifetime. FDID diagrams are conventionally constructed from time-binned data, where it is interpreted as a simple histogram in which each time-bin contributes a single histogram count to one single intensity-decay rate bin. It is not trivial to extend this notion to CPA-segmented data since CPA segments intrinsically have very different time durations instead of having equal width as in conventional time-binning. We propose two modifications to the construction of a FDID as a histogram. First, instead of representing FDID entries as a single binary entry in just one histogram bin (one time segment contributes one count to a single pixel in a FDID), we propose to incorporate the uncertainty in intensity and decay rate that is associated with each time segment. To this end, each segment contributes to the FDID diagram according to a 2D Gaussian function centered at the CPA-segment decay rate and intensity (total counts $C_j$ in segment $j$ divided by segment duration $T_j$), where the width is given by the fit error in the decay rate and the shot noise error in the segment $\sqrt{C_j}/T_j$. If one would apply this logic to regular time-binned data, giving each Gaussian contributor the same integrated weight, one obtains a diagram similar to a regular FDID histogram except that the results is smooth and with less dependency on a chosen histogram bin width. Instead the feature size in FDID represents the actual uncertainties in intensity and rate.

As a second modification, we propose to reconsider the \textit{weights} of the Gaussians --- i.e., the \textit{integrated} contribution to each entry in the FDID. For time-binned data one assigns each segment equal weight so that equal lengths of time contribute equal weight. Since CPA results in segments of unequal length, several choices for constructing FDID diagrams are possible, which to our knowledge have not been discussed in CPA literature. Giving equal weight to each CPA fragment will lead to FDID  diagrams from those obtained from binned data, since effectively long time segments are then underrepresented compared to short segments. This leads to under representation of states with steeper powerlaw distributions in their switching times.  The direct equivalent to regular FDID-weighting for CPA-segmented data is that a segment of duration $T_i$ has a weight proportional to $T_i$ (henceforth `duration-weighted FDIDs'). Alternatively one could argue that since \textit{time-averaged} intensity and fluorescence decay traces are rather set by the contribution in emitted photons, one could instead use the total number of photons $C_i$ in each segment as weight (henceforth `count-weighted FDID').
 
It has been established in a multitude of studies of II-VI quantum dots that the distribution of on/off times follows a power-law (exponents $\alpha_{\mathrm{on, off}}$) truncated by an exponential with specific time $\tau_l$, giving the distribution\cite{Frantsuzov2008,ToddKrauss2010,Efros2016} $t^{-\alpha}e^{-t/\tau_l}$.
We analyze Monte Carlo simulated FDIDs to establish if there are conditions of the truncation time under which a two-state quantum dot would appear not as a bimodal distribution. {Figure~\ref{fig:FDIDs}} shows duration-weighted FDIDs for simulated quantum dot data. For panel (A) we consider a two-state quantum dot with power-law distributed switching times. In  panels (B-D) we show a quantum dot simulated with similar parameters, but with the maximum duration of the segments $T_c = 10, \, 1, \, 0.1$ seconds. Evidently, the bi-modal nature of the quantum dot is faithfully represented by the FDID diagram constructed through CPA. This remains true also for power laws with a long time truncation ($t^{-\alpha}e^{-t/\tau_l}$), unless truncation time $\tau_l$ are as short as 20 ms, so that there are no segments with over approximately 10$^2$ counts. This limit of our benchmarking space zooms in on the regime where CPA intrinsically fails {(Fig.~\ref{fig:sim_likelihoods_detect}(C))}. In this regime, the originally assumed bimodal quantum dot behavior no longer results in a bi-modal FDID. Instead a significant broadening is evident.  We can conclude that for most realistic quantum dot systems, CPA-generated FDIDs will not suffer from this artificial broadening artefact.

\section{Conclusion}
In this work, we have provided a Python toolbox for change-point analysis, and for  determining the most likely intensity level assignment for intermittent multilevel emitters. We 
investigated the limits of changepoint analysis and clustering as fundamentally set by photon budget,  and for the case of many state emitters.  We have shown that for long switching times, the typical power-law behavior of many quantum emitters can be accurately retrieved. We also show that in the case of  many-state emitters, the number of intensity levels can be retrieved with high fidelity, provided the photon count is high enough. At low photon counts, the number of states is systematically underestimated. This shows in which way the photon budget puts a fundamental limit on the amount of information that can be retrieved from a given TCSPC data set. We show that the photon budget also poses a limitation on the accuracy at which the slope of a single-exponential decay can be retrieved.  Additionally, we investigate the commonly used intensity-decay time diagrams. We show that with CPA, a two-state simulated quantum dot is well-represented in an FDID, but when a cutoff is introduced to match that commonly found in literature, the states become increasingly poorly defined in an FDID representation. \textcolor{black}{While the Bayesian inference algorithms in this toolbox were reported earlier for application to quantum dots with just two or three intensity  states,  this toolbox and the provided benchmarks point at the applicability even to emitters that jump between many closely spaced intensity levels,  which  will, in our view, be of large practical use for many workers analyzing the complex photophysics of, e.g., perovskite quantum dots.} Also, the toolbox can be used for theory development, following a workflow in which hypotheses are cast in synthetic photon counting data, which in turn can be subjected to the CPA analysis suite, to assess  how hypothesized mechanisms express in observables, and in how far they are testable. \textcolor{black}{The limit of this testability generally depends on an interplay of total photon budgets,  residence time in each level, intensity contrast between levels,  and segment durations.  For a given photophysics scenario, the user can easily deploy the toolbox to directly assess data segmentation in terms of accuracy, precision and recall error rates, in dependence of the level of skepticism that the user wishes to apply in order to accept assertions regarding the segmentation of data in segments and intensity levels.  These error rates, and hence the testability of a hypothesized photophysics scenario, are ultimately limited by the evidence in the counting statistics, and not the segmentation algorithm.}

\begin{suppinfo}
Detailed manual to the code and description of the example data set. 
\end{suppinfo}

\begin{acknowledgement}
This work is part of the research program of the Netherlands Organization for Scientific Research (NWO). We are grateful to Ilan Shlesinger and Erik Garnett for their critical scrutiny of paper and code.    Lastly we would like to express our gratitude to  Tom Gregorkiewicz who passed away in 2019, and whose encouragement and guidance in the early stages of the project were invaluable.
\end{acknowledgement}
 
%%%%%%%%%%%%%%%%%%%%%%%%%%%%%%%%%%%%%%%%%%%%%%%%%%%
\bibliography{PalstraPython}

\providecommand{\latin}[1]{#1}
\makeatletter
\providecommand{\doi}
  {\begingroup\let\do\@makeother\dospecials
  \catcode`\{=1 \catcode`\}=2 \doi@aux}
\providecommand{\doi@aux}[1]{\endgroup\texttt{#1}}
\makeatother
\providecommand*\mcitethebibliography{\thebibliography}
\csname @ifundefined\endcsname{endmcitethebibliography}
  {\let\endmcitethebibliography\endthebibliography}{}
\begin{mcitethebibliography}{52}
\providecommand*\natexlab[1]{#1}
\providecommand*\mciteSetBstSublistMode[1]{}
\providecommand*\mciteSetBstMaxWidthForm[2]{}
\providecommand*\mciteBstWouldAddEndPuncttrue
  {\def\EndOfBibitem{\unskip.}}
\providecommand*\mciteBstWouldAddEndPunctfalse
  {\let\EndOfBibitem\relax}
\providecommand*\mciteSetBstMidEndSepPunct[3]{}
\providecommand*\mciteSetBstSublistLabelBeginEnd[3]{}
\providecommand*\EndOfBibitem{}
\mciteSetBstSublistMode{f}
\mciteSetBstMaxWidthForm{subitem}{(\alph{mcitesubitemcount})}
\mciteSetBstSublistLabelBeginEnd
  {\mcitemaxwidthsubitemform\space}
  {\relax}
  {\relax}

\bibitem[Orrit and Bernard(1990)Orrit, and Bernard]{Orrit1990}
Orrit,~M.; Bernard,~J. Single pentacene molecules detected by fluorescence
  excitation in a p-terphenyl crystal. \emph{Phys. Rev. Lett.} \textbf{1990},
  \emph{65}, 2716--2719\relax
\mciteBstWouldAddEndPuncttrue
\mciteSetBstMidEndSepPunct{\mcitedefaultmidpunct}
{\mcitedefaultendpunct}{\mcitedefaultseppunct}\relax
\EndOfBibitem
\bibitem[Lounis and Orrit(2005)Lounis, and Orrit]{Lounis2005}
Lounis,~B.; Orrit,~M. Single-photon sources. \emph{Rep. Prog. Phys.}
  \textbf{2005}, \emph{68}, 1129--1179\relax
\mciteBstWouldAddEndPuncttrue
\mciteSetBstMidEndSepPunct{\mcitedefaultmidpunct}
{\mcitedefaultendpunct}{\mcitedefaultseppunct}\relax
\EndOfBibitem
\bibitem[Kimble(2008)]{Kimble2008}
Kimble,~H.~J. The quantum internet. \emph{Nature} \textbf{2008}, \emph{453},
  1023--1030\relax
\mciteBstWouldAddEndPuncttrue
\mciteSetBstMidEndSepPunct{\mcitedefaultmidpunct}
{\mcitedefaultendpunct}{\mcitedefaultseppunct}\relax
\EndOfBibitem
\bibitem[Lodahl \latin{et~al.}(2015)Lodahl, Mahmoodian, and Stobbe]{Lodahl2015}
Lodahl,~P.; Mahmoodian,~S.; Stobbe,~S. Interfacing single photons and single
  quantum dots with photonic nanostructures. \emph{Rev. Mod. Phys.}
  \textbf{2015}, \emph{87}, 347--400\relax
\mciteBstWouldAddEndPuncttrue
\mciteSetBstMidEndSepPunct{\mcitedefaultmidpunct}
{\mcitedefaultendpunct}{\mcitedefaultseppunct}\relax
\EndOfBibitem
\bibitem[Somaschi \latin{et~al.}(2016)Somaschi, Giesz, Santis, Loredo, Almeida,
  Hornecker, Portalupi, Grange, Ant'on, Demory, Gomez, Sagnes, Kimura,
  Lema{\^i}tre, Auff{\`e}ves, White, Lanco, and Senellart]{Somaschi2016}
Somaschi,~N. \latin{et~al.}  Near-optimal single-photon sources in the solid
  state. \emph{Nat. Photonics} \textbf{2016}, \emph{10}, 340--345\relax
\mciteBstWouldAddEndPuncttrue
\mciteSetBstMidEndSepPunct{\mcitedefaultmidpunct}
{\mcitedefaultendpunct}{\mcitedefaultseppunct}\relax
\EndOfBibitem
\bibitem[Doherty \latin{et~al.}(2013)Doherty, Manson, Delaney, Jelezko,
  Wrachtrup, and Hollenberg]{Doherty}
Doherty,~M.~W.; Manson,~N.~B.; Delaney,~P.; Jelezko,~F.; Wrachtrup,~J.;
  Hollenberg,~C.~L. The nitrogen-vacancy colour centre in diamond. \emph{Phys.
  Rep.} \textbf{2013}, \emph{528}, 1 -- 45\relax
\mciteBstWouldAddEndPuncttrue
\mciteSetBstMidEndSepPunct{\mcitedefaultmidpunct}
{\mcitedefaultendpunct}{\mcitedefaultseppunct}\relax
\EndOfBibitem
\bibitem[Castelletto and Boretti(2020)Castelletto, and Boretti]{Castelletto}
Castelletto,~S.; Boretti,~A. Silicon carbide color centers for quantum
  applications. \emph{J. Phys: Photonics} \textbf{2020}, \emph{2}, 022001\relax
\mciteBstWouldAddEndPuncttrue
\mciteSetBstMidEndSepPunct{\mcitedefaultmidpunct}
{\mcitedefaultendpunct}{\mcitedefaultseppunct}\relax
\EndOfBibitem
\bibitem[Toth and Aharonovich(2019)Toth, and Aharonovich]{Aharonovich}
Toth,~M.; Aharonovich,~I. Single Photon Sources in Atomically Thin Materials.
  \emph{Annu. Rev. Phys. Chem.} \textbf{2019}, \emph{70}, 123--142\relax
\mciteBstWouldAddEndPuncttrue
\mciteSetBstMidEndSepPunct{\mcitedefaultmidpunct}
{\mcitedefaultendpunct}{\mcitedefaultseppunct}\relax
\EndOfBibitem
\bibitem[Toninelli \latin{et~al.}(2020)Toninelli, Gerhardt, Clark,
  Reserbat-Plantey, G\"{o}tzinger, Ristanovic, Colautti, Lombardi, Major,
  Deperasi\'{n}ska, Pernice, Koppens, Kozankiewicz, Gourdon, Sandoghdar, and
  Orrit]{Toninelli}
Toninelli,~C. \latin{et~al.}  Single organic molecules for photonic quantum
  technologies. \emph{preprint} \textbf{2020}, arXiv:2011.05059\relax
\mciteBstWouldAddEndPuncttrue
\mciteSetBstMidEndSepPunct{\mcitedefaultmidpunct}
{\mcitedefaultendpunct}{\mcitedefaultseppunct}\relax
\EndOfBibitem
\bibitem[Murray \latin{et~al.}(1993)Murray, Norris, and Bawendi]{Murray1993}
Murray,~C.~B.; Norris,~D.~J.; Bawendi,~M.~G. Synthesis and characterization of
  nearly monodisperse CdE (E = sulfur, selenium, tellurium) semiconductor
  nanocrystallites. \emph{J. Am. Chem. Soc.} \textbf{1993}, \emph{115},
  8706--8715\relax
\mciteBstWouldAddEndPuncttrue
\mciteSetBstMidEndSepPunct{\mcitedefaultmidpunct}
{\mcitedefaultendpunct}{\mcitedefaultseppunct}\relax
\EndOfBibitem
\bibitem[Talapin \latin{et~al.}(2010)Talapin, Lee, Kovalenko, and
  Shevchenko]{Talapin}
Talapin,~D.~V.; Lee,~J.-S.; Kovalenko,~M.~V.; Shevchenko,~E.~V. Prospects of
  Colloidal Nanocrystals for Electronic and Optoelectronic Applications.
  \emph{Chem. Rev.} \textbf{2010}, \emph{110}, 389--458\relax
\mciteBstWouldAddEndPuncttrue
\mciteSetBstMidEndSepPunct{\mcitedefaultmidpunct}
{\mcitedefaultendpunct}{\mcitedefaultseppunct}\relax
\EndOfBibitem
\bibitem[Shirasaki \latin{et~al.}(2013)Shirasaki, Supran, Bawendi, and
  Bulovic]{Shirasaki}
Shirasaki,~Y.; Supran,~G.; Bawendi,~M.; Bulovic,~V. Emergence of colloidal
  quantum-dot light-emitting technologies. \emph{Nat. Photonics} \textbf{2013},
  \emph{7}, 13--23\relax
\mciteBstWouldAddEndPuncttrue
\mciteSetBstMidEndSepPunct{\mcitedefaultmidpunct}
{\mcitedefaultendpunct}{\mcitedefaultseppunct}\relax
\EndOfBibitem
\bibitem[Protesescu \latin{et~al.}(2015)Protesescu, Yakunin, Bodnarchuk, Krieg,
  Caputo, Hendon, Yang, Walsh, and Kovalenko]{Protesescu2015}
Protesescu,~L.; Yakunin,~S.; Bodnarchuk,~M.~I.; Krieg,~F.; Caputo,~R.;
  Hendon,~C.~H.; Yang,~R.~X.; Walsh,~A.; Kovalenko,~M.~V. {Nanocrystals of
  Cesium Lead Halide Perovskites (CsPbX 3 , X = Cl, Br, and I): Novel
  Optoelectronic Materials Showing Bright Emission with Wide Color Gamut}.
  \emph{Nano Lett.} \textbf{2015}, \emph{15}, 3692--3696\relax
\mciteBstWouldAddEndPuncttrue
\mciteSetBstMidEndSepPunct{\mcitedefaultmidpunct}
{\mcitedefaultendpunct}{\mcitedefaultseppunct}\relax
\EndOfBibitem
\bibitem[Swarnkar \latin{et~al.}(2015)Swarnkar, Chulliyil, Ravi, Irfanullah,
  Chowdhury, and Nag]{Swarnkar2015}
Swarnkar,~A.; Chulliyil,~R.; Ravi,~V.; Irfanullah,~M.; Chowdhury,~A.; Nag,~A.
  Colloidal CsPbBr 3 Perovskite Nanocrystals: Luminescence beyond Traditional
  Quantum Dots. \emph{Angew. Chem. Int. Ed.} \textbf{2015}, \relax
\mciteBstWouldAddEndPunctfalse
\mciteSetBstMidEndSepPunct{\mcitedefaultmidpunct}
{}{\mcitedefaultseppunct}\relax
\EndOfBibitem
\bibitem[Park \latin{et~al.}(2015)Park, Guo, Makarov, and Klimov]{Park2015}
Park,~Y.-S.; Guo,~S.; Makarov,~N.~S.; Klimov,~V.~I. Room Temperature
  Single-Photon Emission from Individual Perovskite Quantum Dots. \emph{ACS
  Nano} \textbf{2015}, \emph{9}, 10386--10393\relax
\mciteBstWouldAddEndPuncttrue
\mciteSetBstMidEndSepPunct{\mcitedefaultmidpunct}
{\mcitedefaultendpunct}{\mcitedefaultseppunct}\relax
\EndOfBibitem
\bibitem[Li \latin{et~al.}(2018)Li, Huang, Zhang, Yang, Guo, Chen, Qin, Gao,
  Biju, Rogach, Xiao, and Jia]{LiHuang2018}
Li,~B.; Huang,~H.; Zhang,~G.; Yang,~C.; Guo,~W.; Chen,~R.; Qin,~C.; Gao,~Y.;
  Biju,~V.~P.; Rogach,~A.~L.; Xiao,~L.; Jia,~S. Excitons and Biexciton Dynamics
  in Single CsPbBr$_3$ Perovskite Quantum Dots. \emph{J. Phys. Chem. Lett.}
  \textbf{2018}, \emph{9}, 6934--6940\relax
\mciteBstWouldAddEndPuncttrue
\mciteSetBstMidEndSepPunct{\mcitedefaultmidpunct}
{\mcitedefaultendpunct}{\mcitedefaultseppunct}\relax
\EndOfBibitem
\bibitem[Gibson \latin{et~al.}(2018)Gibson, Koscher, Alivisatos, and
  Leone]{Gibson2018}
Gibson,~N.~A.; Koscher,~B.~A.; Alivisatos,~A.~P.; Leone,~S.~R. {Excitation
  Intensity Dependence of Photoluminescence Blinking in CsPbBr$_3$ Perovskite
  Nanocrystals}. \emph{J. Phys. Chem. C} \textbf{2018}, \emph{122},
  12106--12113\relax
\mciteBstWouldAddEndPuncttrue
\mciteSetBstMidEndSepPunct{\mcitedefaultmidpunct}
{\mcitedefaultendpunct}{\mcitedefaultseppunct}\relax
\EndOfBibitem
\bibitem[Seth \latin{et~al.}(2016)Seth, Mondal, Patra, and Samanta]{Seth2016}
Seth,~S.; Mondal,~N.; Patra,~S.; Samanta,~A. Fluorescence Blinking and
  Photoactivation of All-Inorganic Perovskite Nanocrystals CsPbBr$_3$ and
  CsPbBr$_2$I. \emph{J. Phys. Chem. Lett.} \textbf{2016}, \emph{7},
  266--271\relax
\mciteBstWouldAddEndPuncttrue
\mciteSetBstMidEndSepPunct{\mcitedefaultmidpunct}
{\mcitedefaultendpunct}{\mcitedefaultseppunct}\relax
\EndOfBibitem
\bibitem[Yuan \latin{et~al.}(2018)Yuan, Ritchie, Ritter, Murphy, Gómez, and
  Mulvaney]{Yuan2018}
Yuan,~G.; Ritchie,~C.; Ritter,~M.; Murphy,~S.; Gómez,~D.~E.; Mulvaney,~P. The
  Degradation and Blinking of Single CsPbI$_3$ Perovskite Quantum Dots.
  \emph{J. Phys. Chem. C} \textbf{2018}, \emph{122}, 13407--13415\relax
\mciteBstWouldAddEndPuncttrue
\mciteSetBstMidEndSepPunct{\mcitedefaultmidpunct}
{\mcitedefaultendpunct}{\mcitedefaultseppunct}\relax
\EndOfBibitem
\bibitem[Hou \latin{et~al.}(2020)Hou, Zhao, Yuan, Zhao, Krieg, Tamarat,
  Kovalenko, Guo, and Lounis]{HouKovalenko2020}
Hou,~L.; Zhao,~C.; Yuan,~X.; Zhao,~J.; Krieg,~F.; Tamarat,~P.;
  Kovalenko,~M.~V.; Guo,~C.; Lounis,~B. Memories in the photoluminescence
  intermittency of single cesium lead bromide nanocrystals. \emph{Nanoscale}
  \textbf{2020}, \emph{12}, 6795--6802\relax
\mciteBstWouldAddEndPuncttrue
\mciteSetBstMidEndSepPunct{\mcitedefaultmidpunct}
{\mcitedefaultendpunct}{\mcitedefaultseppunct}\relax
\EndOfBibitem
\bibitem[Frantsuzov \latin{et~al.}(2008)Frantsuzov, Kuno, Jank\'{o}, and
  Marcus]{Frantsuzov2008}
Frantsuzov,~P.; Kuno,~M.; Jank\'{o},~B.; Marcus,~R. Universal emission
  intermittency in quantum dots, nanorods, and nanowires. \emph{Nat. Phys.}
  \textbf{2008}, \emph{4}, 519\relax
\mciteBstWouldAddEndPuncttrue
\mciteSetBstMidEndSepPunct{\mcitedefaultmidpunct}
{\mcitedefaultendpunct}{\mcitedefaultseppunct}\relax
\EndOfBibitem
\bibitem[Krauss and Peterson(2010)Krauss, and Peterson]{ToddKrauss2010}
Krauss,~T.~D.; Peterson,~J.~J. Bright Future for Fluorescence Blinking in
  Semiconductor Nanocrystals. \emph{J. Phys. Chem. Lett.} \textbf{2010},
  \emph{1}, 1377--1382\relax
\mciteBstWouldAddEndPuncttrue
\mciteSetBstMidEndSepPunct{\mcitedefaultmidpunct}
{\mcitedefaultendpunct}{\mcitedefaultseppunct}\relax
\EndOfBibitem
\bibitem[Efros and Nesbitt(2016)Efros, and Nesbitt]{Efros2016}
Efros,~A.; Nesbitt,~D. Origin and control of blinking in quantum dots.
  \emph{Nat. Nanotechn.} \textbf{2016}, \emph{11}, 661--671\relax
\mciteBstWouldAddEndPuncttrue
\mciteSetBstMidEndSepPunct{\mcitedefaultmidpunct}
{\mcitedefaultendpunct}{\mcitedefaultseppunct}\relax
\EndOfBibitem
\bibitem[Galland \latin{et~al.}(2011)Galland, Ghosh, Steinbr{\"{u}}ck, Sykora,
  Hollingsworth, Klimov, and Htoon]{Galland2011}
Galland,~C.; Ghosh,~Y.; Steinbr{\"{u}}ck,~A.; Sykora,~M.; Hollingsworth,~J.~A.;
  Klimov,~V.~I.; Htoon,~H. {Two types of luminescence blinking revealed by
  spectroelectrochemistry of single quantum dots}. \emph{Nature} \textbf{2011},
  \emph{479}, 203--207\relax
\mciteBstWouldAddEndPuncttrue
\mciteSetBstMidEndSepPunct{\mcitedefaultmidpunct}
{\mcitedefaultendpunct}{\mcitedefaultseppunct}\relax
\EndOfBibitem
\bibitem[Rabouw \latin{et~al.}(2013)Rabouw, Lunnemann, van Dijk-Moes, Frimmer,
  Pietra, Koenderink, and Vanmaekelbergh]{Rabouw2013}
Rabouw,~F.~T.; Lunnemann,~P.; van Dijk-Moes,~R. J.~A.; Frimmer,~M.; Pietra,~F.;
  Koenderink,~A.~F.; Vanmaekelbergh,~D. {Reduced Auger Recombination in Single
  CdSe/CdS Nanorods by One-Dimensional Electron Delocalization}. \emph{Nano
  Lett.} \textbf{2013}, \emph{13}, 4884--4892\relax
\mciteBstWouldAddEndPuncttrue
\mciteSetBstMidEndSepPunct{\mcitedefaultmidpunct}
{\mcitedefaultendpunct}{\mcitedefaultseppunct}\relax
\EndOfBibitem
\bibitem[Watkins and Yang(2005)Watkins, and Yang]{Watkins2005}
Watkins,~L.~P.; Yang,~H. {Detection of Intensity Change Points in Time-Resolved
  Single-Molecule Measurements}. \emph{J. Phys. Chem. B} \textbf{2005},
  \emph{109}, 617--628\relax
\mciteBstWouldAddEndPuncttrue
\mciteSetBstMidEndSepPunct{\mcitedefaultmidpunct}
{\mcitedefaultendpunct}{\mcitedefaultseppunct}\relax
\EndOfBibitem
\bibitem[Hoogenboom \latin{et~al.}(2006)Hoogenboom, den Otter, and
  Offerhaus]{Hoogenboom2006}
Hoogenboom,~J.~P.; den Otter,~W.~K.; Offerhaus,~H.~L. {Accurate and unbiased
  estimation of power-law exponents from single-emitter blinking data}.
  \emph{J. Chem. Phys} \textbf{2006}, \emph{125}, 204713\relax
\mciteBstWouldAddEndPuncttrue
\mciteSetBstMidEndSepPunct{\mcitedefaultmidpunct}
{\mcitedefaultendpunct}{\mcitedefaultseppunct}\relax
\EndOfBibitem
\bibitem[Ensign and Pande(2009)Ensign, and Pande]{Ensign2009}
Ensign,~D.~L.; Pande,~V.~S. {Bayesian Single-Exponential Kinetics in
  Single-Molecule Experiments and Simulations}. \emph{J. Phys. Chem. B}
  \textbf{2009}, \emph{113}, 12410--12423\relax
\mciteBstWouldAddEndPuncttrue
\mciteSetBstMidEndSepPunct{\mcitedefaultmidpunct}
{\mcitedefaultendpunct}{\mcitedefaultseppunct}\relax
\EndOfBibitem
\bibitem[Ensign and Pande(2010)Ensign, and Pande]{Ensign2010}
Ensign,~D.~L.; Pande,~V.~S. Bayesian Detection of Intensity Changes in Single
  Molecule and Molecular Dynamics Trajectories. \emph{J. Phys. Chem. B}
  \textbf{2010}, \emph{114}, 280--292\relax
\mciteBstWouldAddEndPuncttrue
\mciteSetBstMidEndSepPunct{\mcitedefaultmidpunct}
{\mcitedefaultendpunct}{\mcitedefaultseppunct}\relax
\EndOfBibitem
\bibitem[Crouch \latin{et~al.}(2010)Crouch, Sauter, Wu, Purcell, Querner,
  Drndic, and Pelton]{Crouch2010}
Crouch,~C.~H.; Sauter,~O.; Wu,~X.; Purcell,~R.; Querner,~C.; Drndic,~M.;
  Pelton,~M. {Facts and Artifacts in the Blinking Statistics of Semiconductor
  Nanocrystals}. \emph{Nano Lett.} \textbf{2010}, \emph{10}, 1692--1698\relax
\mciteBstWouldAddEndPuncttrue
\mciteSetBstMidEndSepPunct{\mcitedefaultmidpunct}
{\mcitedefaultendpunct}{\mcitedefaultseppunct}\relax
\EndOfBibitem
\bibitem[Houel \latin{et~al.}(2015)Houel, Doan, Cajgfinger, Ledoux, Amans,
  Aubret, Dominjon, Ferriol, Barbier, Nasilowski, Lhuillier, Dubertret,
  Dujardin, and Kulzer]{Houel2015}
Houel,~J.; Doan,~Q.~T.; Cajgfinger,~T.; Ledoux,~G.; Amans,~D.; Aubret,~A.;
  Dominjon,~A.; Ferriol,~S.; Barbier,~R.; Nasilowski,~M.; Lhuillier,~E.;
  Dubertret,~B.; Dujardin,~C.; Kulzer,~F. Autocorrelation Analysis for the
  Unbiased Determination of Power-Law Exponents in Single-Quantum-Dot Blinking.
  \emph{ACS Nano} \textbf{2015}, \emph{9}, 886--893\relax
\mciteBstWouldAddEndPuncttrue
\mciteSetBstMidEndSepPunct{\mcitedefaultmidpunct}
{\mcitedefaultendpunct}{\mcitedefaultseppunct}\relax
\EndOfBibitem
\bibitem[Bae \latin{et~al.}(2016)Bae, Gibson, Ding, Alivisatos, and
  Leone]{Bae2016}
Bae,~Y.~J.; Gibson,~N.~A.; Ding,~T.~X.; Alivisatos,~A.~P.; Leone,~S.~R.
  {Understanding the Bias Introduced in Quantum Dot Blinking Using Change Point
  Analysis}. \emph{J. Phys. Chem. C} \textbf{2016}, \emph{120},
  29484--29490\relax
\mciteBstWouldAddEndPuncttrue
\mciteSetBstMidEndSepPunct{\mcitedefaultmidpunct}
{\mcitedefaultendpunct}{\mcitedefaultseppunct}\relax
\EndOfBibitem
\bibitem[Hill \latin{et~al.}(2017)Hill, Monachino, and van Oijen]{Hill}
Hill,~F.~R.; Monachino,~E.; van Oijen,~A.~M. {The more the merrier:
  high-throughput single-molecule techniques}. \emph{Biochem. Soc. Trans.}
  \textbf{2017}, \emph{45}, 759--769\relax
\mciteBstWouldAddEndPuncttrue
\mciteSetBstMidEndSepPunct{\mcitedefaultmidpunct}
{\mcitedefaultendpunct}{\mcitedefaultseppunct}\relax
\EndOfBibitem
\bibitem[Hadzic \latin{et~al.}(2018)Hadzic, B\"{o}rner, K\"{o}nig, Kowerko, and
  Sigel]{Hadzic}
Hadzic,~M. C. A.~S.; B\"{o}rner,~R.; K\"{o}nig,~S. L.~B.; Kowerko,~D.;
  Sigel,~R. K.~O. Reliable State Identification and State Transition Detection
  in Fluorescence Intensity-Based Single-Molecule Förster Resonance
  Energy-Transfer Data. \emph{J. Phys. Chem. B} \textbf{2018}, \emph{122},
  6134--6147\relax
\mciteBstWouldAddEndPuncttrue
\mciteSetBstMidEndSepPunct{\mcitedefaultmidpunct}
{\mcitedefaultendpunct}{\mcitedefaultseppunct}\relax
\EndOfBibitem
\bibitem[Schmid \latin{et~al.}(2016)Schmid, G\"{o}tz, and Hugel]{schmid2016}
Schmid,~S.; G\"{o}tz,~M.; Hugel,~T. Single-Molecule Analysis beyond Dwell
  Times: Demonstration and Assessment in and out of Equilibrium.
  \emph{Biophysical Journal} \textbf{2016}, \emph{111}, 1375--1384\relax
\mciteBstWouldAddEndPuncttrue
\mciteSetBstMidEndSepPunct{\mcitedefaultmidpunct}
{\mcitedefaultendpunct}{\mcitedefaultseppunct}\relax
\EndOfBibitem
\bibitem[Press\'{e} \latin{et~al.}(2014)Press\'{e}, Peterson, Lee, Elms,
  MacCallum, Marqusee, Bustamante, and Dill]{Presse2014}
Press\'{e},~S.; Peterson,~J.; Lee,~J.; Elms,~P.; MacCallum,~J.~L.;
  Marqusee,~S.; Bustamante,~C.; Dill,~K. Single Molecule Conformational Memory
  Extraction: P5ab RNA Hairpin. \emph{The Journal of Physical Chemistry B}
  \textbf{2014}, \emph{118}, 6597--6603\relax
\mciteBstWouldAddEndPuncttrue
\mciteSetBstMidEndSepPunct{\mcitedefaultmidpunct}
{\mcitedefaultendpunct}{\mcitedefaultseppunct}\relax
\EndOfBibitem
\bibitem[White \latin{et~al.}(2020)White, Goldschen-Ohm, Goldsmith, and
  Chanda]{disc2020}
White,~D.~S.; Goldschen-Ohm,~M.~P.; Goldsmith,~R.~H.; Chanda,~B. Top-down
  machine learning approach for high-throughput single-molecule analysis.
  \emph{eLife} \textbf{2020}, \emph{9}, e53357\relax
\mciteBstWouldAddEndPuncttrue
\mciteSetBstMidEndSepPunct{\mcitedefaultmidpunct}
{\mcitedefaultendpunct}{\mcitedefaultseppunct}\relax
\EndOfBibitem
\bibitem[Shuang \latin{et~al.}(2014)Shuang, Cooper, Taylor, Kisley, Chen, Wang,
  Li, Komatsuzaki, and Landes]{Shuang2014}
Shuang,~B.; Cooper,~D.; Taylor,~J.~N.; Kisley,~L.; Chen,~J.; Wang,~W.;
  Li,~C.~B.; Komatsuzaki,~T.; Landes,~C.~F. Fast Step Transition and State
  Identification (STaSI) for Discrete Single-Molecule Data Analysis. \emph{The
  Journal of Physical Chemistry Letters} \textbf{2014}, \emph{5},
  3157--3161\relax
\mciteBstWouldAddEndPuncttrue
\mciteSetBstMidEndSepPunct{\mcitedefaultmidpunct}
{\mcitedefaultendpunct}{\mcitedefaultseppunct}\relax
\EndOfBibitem
\bibitem[Li and Yang(2019)Li, and Yang]{LiYang2019}
Li,~H.; Yang,~H. Statistical Learning of Discrete States in Time Series.
  \emph{J. Phys. Chem. B} \textbf{2019}, \emph{123}, 689--701\relax
\mciteBstWouldAddEndPuncttrue
\mciteSetBstMidEndSepPunct{\mcitedefaultmidpunct}
{\mcitedefaultendpunct}{\mcitedefaultseppunct}\relax
\EndOfBibitem
\bibitem[Ensign(2010)]{EnsignThesis2010}
Ensign,~D.~L. Bayesian statistics and single-molecule trajectories. Ph.D.\
  thesis, Stanford University, 2010\relax
\mciteBstWouldAddEndPuncttrue
\mciteSetBstMidEndSepPunct{\mcitedefaultmidpunct}
{\mcitedefaultendpunct}{\mcitedefaultseppunct}\relax
\EndOfBibitem
\bibitem[Schmidt \latin{et~al.}(2012)Schmidt, Krasselt, and {Von
  Borczyskowski}]{Schmidt2012}
Schmidt,~R.; Krasselt,~C.; {Von Borczyskowski},~C. {Change point analysis of
  matrix dependent photoluminescence intermittency of single CdSe/ZnS quantum
  dots with intermediate intensity levels}. \emph{Chem. Phys.} \textbf{2012},
  \emph{406}, 9--14\relax
\mciteBstWouldAddEndPuncttrue
\mciteSetBstMidEndSepPunct{\mcitedefaultmidpunct}
{\mcitedefaultendpunct}{\mcitedefaultseppunct}\relax
\EndOfBibitem
\bibitem[Palstra and Koenderink(2021)Palstra, and Koenderink]{github}
Palstra,~I.~M.; Koenderink,~A.~F. \emph{A Python toolbox for unbiased
  statistical analysis of fluorescence intermittency of multi-level emitters.
  GitHub repository \url{https://github.com/AMOLFResonantNanophotonics/CPA/}.
  First commit archived at \url{DOI: 10.5281/zenodo.4557226}}; 2021\relax
\mciteBstWouldAddEndPuncttrue
\mciteSetBstMidEndSepPunct{\mcitedefaultmidpunct}
{\mcitedefaultendpunct}{\mcitedefaultseppunct}\relax
\EndOfBibitem
\bibitem[Palstra \latin{et~al.}(2021)Palstra, Wenniger, Patra, Garnett, and
  Koenderink]{Palstra2021expt}
Palstra,~I.; Wenniger,~I.; Patra,~B.~K.; Garnett,~E.~C.; Koenderink,~A.~F.
  Intermittency of CsPbBr$_3$ perovskite quantum dots analyzed by an unbiased
  statistical analysis. \emph{Submitted J. Phys. Chem. C} \textbf{2021}, Arxiv
  preprint 2102.09333.\relax
\mciteBstWouldAddEndPunctfalse
\mciteSetBstMidEndSepPunct{\mcitedefaultmidpunct}
{}{\mcitedefaultseppunct}\relax
\EndOfBibitem
\bibitem[Zhang \latin{et~al.}(2006)Zhang, Chang, Fu, Alivisatos, and
  Yang]{Zhang2006}
Zhang,~K.; Chang,~H.; Fu,~A.; Alivisatos,~A.~P.; Yang,~H. {Continuous
  Distribution of Emission States from Single CdSe/ZnS Quantum Dots}.
  \emph{Nano Lett.} \textbf{2006}, \emph{6}, 843--847\relax
\mciteBstWouldAddEndPuncttrue
\mciteSetBstMidEndSepPunct{\mcitedefaultmidpunct}
{\mcitedefaultendpunct}{\mcitedefaultseppunct}\relax
\EndOfBibitem
\bibitem[G'{o}mez \latin{et~al.}(2009)G'{o}mez, van Embden, Mulvaney,
  Fern\'{e}e, and Rubinsztein-Dunlop]{Rubinsztein2009}
G'{o}mez,~D.~E.; van Embden,~J.; Mulvaney,~P.; Fern\'{e}e,~M.~J.;
  Rubinsztein-Dunlop,~H. Exciton-Trion Transitions in Single CdSe-CdS
  Core–Shell Nanocrystals. \emph{ACS Nano} \textbf{2009}, \emph{3},
  2281--2287\relax
\mciteBstWouldAddEndPuncttrue
\mciteSetBstMidEndSepPunct{\mcitedefaultmidpunct}
{\mcitedefaultendpunct}{\mcitedefaultseppunct}\relax
\EndOfBibitem
\bibitem[Cordones \latin{et~al.}(2011)Cordones, Bixby, and Leone]{Cordones2011}
Cordones,~A.~A.; Bixby,~T.~J.; Leone,~S.~R. Direct Measurement of Off-State
  Trapping Rate Fluctuations in Single Quantum Dot Fluorescence. \emph{Nano
  Lett.} \textbf{2011}, \emph{11}, 3366--3369\relax
\mciteBstWouldAddEndPuncttrue
\mciteSetBstMidEndSepPunct{\mcitedefaultmidpunct}
{\mcitedefaultendpunct}{\mcitedefaultseppunct}\relax
\EndOfBibitem
\bibitem[Schmidt \latin{et~al.}(2014)Schmidt, Krasselt, Göhler, and von
  Borczyskowski]{Schmidt2014}
Schmidt,~R.; Krasselt,~C.; Göhler,~C.; von Borczyskowski,~C. The Fluorescence
  Intermittency for Quantum Dots Is Not Power-Law Distributed: A Luminescence
  Intensity Resolved Approach. \emph{ACS Nano} \textbf{2014}, \emph{8},
  3506--3521\relax
\mciteBstWouldAddEndPuncttrue
\mciteSetBstMidEndSepPunct{\mcitedefaultmidpunct}
{\mcitedefaultendpunct}{\mcitedefaultseppunct}\relax
\EndOfBibitem
\bibitem[Rabouw \latin{et~al.}(2019)Rabouw, Antolinez, Brechb{\"u}hler, and
  Norris]{Rabouw2019}
Rabouw,~F.~T.; Antolinez,~F.~V.; Brechb{\"u}hler,~R.; Norris,~D.~J. Microsecond
  Blinking Events in the Fluorescence of Colloidal Quantum Dots Revealed by
  Correlation Analysis on Preselected Photons. \emph{J. Phys. Chem. Lett.}
  \textbf{2019}, \emph{10}, 3732 -- 3738\relax
\mciteBstWouldAddEndPuncttrue
\mciteSetBstMidEndSepPunct{\mcitedefaultmidpunct}
{\mcitedefaultendpunct}{\mcitedefaultseppunct}\relax
\EndOfBibitem
\bibitem[Wahl \latin{et~al.}(2003)Wahl, Gregor, Patting, and
  Enderlein]{Wahl2003}
Wahl,~M.; Gregor,~I.; Patting,~M.; Enderlein,~J. {Fast calculation of
  fluorescence correlation data with asynchronous time-correlated single-photon
  counting}. \emph{Optics Express} \textbf{2003}, \emph{11}, 3583\relax
\mciteBstWouldAddEndPuncttrue
\mciteSetBstMidEndSepPunct{\mcitedefaultmidpunct}
{\mcitedefaultendpunct}{\mcitedefaultseppunct}\relax
\EndOfBibitem
\bibitem[Bajzer \latin{et~al.}(1991)Bajzer, Therneau, Sharp, and
  Prendergast]{Bajzer1991}
Bajzer,~{\v{Z}}.; Therneau,~T.~M.; Sharp,~J.~C.; Prendergast,~F.~G. {Maximum
  likelihood method for the analysis of time-resolved fluorescence decay
  curves}. \emph{Eur. Biophys. J.} \textbf{1991}, \emph{20}, 247--262\relax
\mciteBstWouldAddEndPuncttrue
\mciteSetBstMidEndSepPunct{\mcitedefaultmidpunct}
{\mcitedefaultendpunct}{\mcitedefaultseppunct}\relax
\EndOfBibitem
\bibitem[K{\"{o}}llner and Wolfrum(1992)K{\"{o}}llner, and
  Wolfrum]{Kollner1992}
K{\"{o}}llner,~M.; Wolfrum,~J. {How many photons are necessary for
  fluorescence-lifetime measurements ?} \emph{Chem. Phys. Lett.} \textbf{1992},
  \emph{200}, 199--204\relax
\mciteBstWouldAddEndPuncttrue
\mciteSetBstMidEndSepPunct{\mcitedefaultmidpunct}
{\mcitedefaultendpunct}{\mcitedefaultseppunct}\relax
\EndOfBibitem
\end{mcitethebibliography}

\newcommand{\py}[1]{\textsf{#1}}
\renewcommand{\nG}{n_\textrm{G}}
\newcolumntype{L}{>{$}l<{$}} % math-mode version of "l" column type
\newcommand{\dtau}{$d\tau$}

\section{Supporting information - manual to the Python code}
\singlespacing

In this supplement, we will discuss the different uses of the Python toolbox for unbiased statistical analysis. The toolkit accepts both data taken in a TCSPC measurement and data created in Monte-Carlo (MC) simulations.  The code, and this manual  are. We will first discuss the data processing of photon time stamp files, and subsequently the provided routine to simulate $m_0$-level single photon emitters.

\section{Processing toolbox - overview}

\begin{figure*}
  \includegraphics[width=1.0\linewidth]{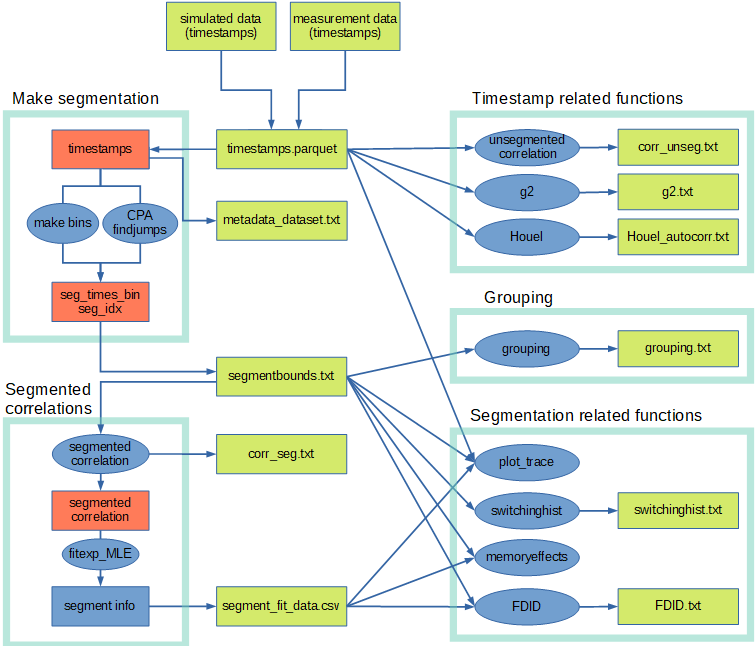}
  \caption{Flow chart of the processing toolbox described in the main text.}\label{fig:flowchart}
\end{figure*}
Figure~\ref{fig:flowchart} shows the different parts of the data analysis tools provided  in the toolbox, and describes a workflow to perform a full analysis. The diagram shows main functions as dark blue ellipses,  input and output files as green boxes,  with  blue frames indicating  groupings or related tasks  the code performs. 

The starting point for the code is a  provided set of three lists of timestamps, which represent three channels,  i.e., two photon counting channels, and a pulsed laser reference record.  These timestamps are provided as sorted lists of 64-bit integers through "parquet" files. To convert integer timestamps to actual times, the user can specify  a discretisation unit, which for experiments would generally be the intrinsic binsize of the TCSPC card used. The data is expected to be in "reverse start-stop" format as usual for TCSPC experiment, meaning that the pulsed laser reference trace contains time stamps for the laser pulses immediately after each photon event.  Simulated data is simulated with a user specified intrinsic binsize and assumed pulsed excitation repetition rate.

The processing code has the following functionality.  First,  the data is segmented into intervals using changepoint analysis.  This returns the jumptimes at which the emitter at hand jumps in intensity, and between which the dot is constant in intensity, within the CPA threshold.  Next, for each segment, the detector and photon channels are time correlated (block "segmented correlations")  to obtain fluorescence decay trace photon counting histograms per segment, to which a decay rate can be fitted with the maximum likelyhood method (MLE).  These processes lead to a set of intermediate, ASCII-formatted files that contain the segmented jumptimes, intensities and decay rates.  The reminder of the routines are found on the right hand side of the diagram. The top block "timestamp related functions" acts not on the segments, but on the timestamp list as a whole. They convert the timestamps list into   time-correlations between detectors, and detector and laser, across the whole dataset, to provide a decay trace, $g^{(2)(\tau)}$ and long-term autocorrelation trace.  The remainder of the blocks post-processes the segmented data. This includes on one hand \textit{grouping} according to Bayesian analysis, and on the other hand  plotting of the segmented data as time traces, FDID diagrams, and switching time histograms.

The philosophy of the code is that it can interchangeably run on experimental and simulated data simply by setting a Boolean switch in the code. The Boolean switch redirects the routine to take information from the \py{preamble\_XXX.py} files, where \py{XXX} points to  experimental resp. simulated data. The preamble files contain the input and output data file paths and required metadata for processing.  The input time stamps are taken from the folders labelled \py{Data\_XXX},  while the output is saved in the folder \py{Output\_XXX} with  Each function can be run independently provided that any required  intermediate files have already been generated.   

To demonstrate a complete run through the entire workflow, we provide the script  \py{EntireWorkflowScript.py}.  This script is just a sequence of independent function calls, with no passing of variables between functions. We note that to work with experimental example, the user needs only run or adapt \py{EntireWorkflowScript.py}.  For exploring a large variety of simulated problems, it suffices to run the simulation routine (see below), and work with \py{EntireWorkflowScript.py}, largely without need to adapt \py{preamble\_simulated.py}.  

Under the hood,  the actual dataprocessing algorithms are contained in the four basic routines:
\begin{enumerate}
    \item  \py{CPA\_functions.py}  --- segments the timestamp file using changepoint analysis
    \item  \py{grouping\_functions.py}  --- clusters lists of jumptimes and counts per segment into the most likely set of levels, according to Bayesian information criterion. 
    \item  \py{correlate\_jit.py}  ---  contains the timestamp-based  channel auto and cross-correlation algorithm reported by  Wahl et al.~\cite{Wahl2003}. It provides decay trace histograms when correlating photon  and laser channel,  used for both segments ("segmented correlations") and aggregate data record ("unsegmented correlation").  It provides $g^{(2)}(\tau)$, when giving  detector  channels A and B as task to correlate,  and it provides on a logarithmically coarsened time axis the normalized intensity autocorrelation to perform the intermittency analysis proposed by Houel et al~\cite{Houel2015} . We note that for typical time records of $10^7-10^8 $ events per channel, obtaining the $g^(2)(\tau)$ photon-photon correlations are the most time-intensive.  For this reason this part of the code is accelerated with the Python \py{numba jit} just in time compiler, and only called with 64-bit integer datatypes.
    \item  \py{fitexp\_MLE.py} --- performs maximum likelihood estimation based fitting of fluorescence decay traces according to Bajzer et al.~\cite{Bajzer1991}, which is applied to both segmented correlations and the aggregate time trace.
\end{enumerate}
The remainder of the routines are for file input-output,  bookkeeping and plotting, relying on the core routines listed above for the actual work.

%%%%%%%%%%%%%%%%%%%%%%%%%%%%%%%%%%%%%%%%%%%%%%%%%%%%%%%%%%%%%%%%%%%%%%%%%
\subsection{Segmenting the data}
\label{SIsec:segmenting}
Module name: \py{CPA\_wrapper.py}\\
Input files: \py{timestamps\_chX\_bin.parquet}\\
Output files: \py{metadata\_dataset.txt}, \py{segmentbounds.txt}\\

In this script, the data  --- experimental or generated by the dot-simulation routine ---  is segmented according to the user's instructions.  The script loads the data from three parquet files, which are binary data files containing all the photon event timestamps on the three channels: the two measurement channels from the Hanbury-Brown Twiss setup of photo-detectors, and the reference channel. This is a ubiquitous setup for single-photon detection, but we note that, in the case of a different number of detectors, the code can be adapted to accommodate for this, as in most scripts (except for $g(2)(\tau)$), the events from the two photodetectors are merged into a single list.

The script initially retrieves a number of properties about the data, such as  the number of counts in all channels, and the length of the time record (timing of the last events),  which together with the timing bin width (specified in \py{preamble.py}) are written to a metadata file. This file also includes a retrieve time offset between channels A, B and R estimated from the data. That is, an event on a detector and the corresponding reference event will be offset by (1) the time it takes for the emitter to decay, and (2) the path difference between detectors and reference diode. The quantities \py{Tau0A\_bin, Tau0B\_bin} refer to the latter.
In simulated data there is no such time offset, but in experimental data there are typically optical and electronic (cable length) path differences which one needs to correct for before splicing two detector channels into a single fluorescence decay trace.  

The script next calls \py{cpa\_functions.py}, which takes the combined photon timestamps from channel A and B as input, and returns the timestamps at which a  changepoint occurs according to Bayesian inference. 
This routine takes the basic formalism  to find a single changepoint in a dataset that has photon timestamps distributed according to exponential waiting time statistics, as specified by Watkins~\cite{Watkins2005}, Ensign~\cite{EnsignThesis2010}, and in the main text. This formalism is applied iteratively to the resulting segments, subdividing them until no more changepoints are found. 
The routines returns both the timestamps (jump times $t_0, t_1$, $t_2$, ...) and the associated photon event indices, which is equivalent to the cumulative number of counts up to the time of the jump.  The single-changepoint detection routine requires a level of `skepticism' to accept or reject changepoints, set in the preamble python file. For an explanation on the which values are appropriate, we point the reader to~\cite{Watkins2005} and~\cite{EnsignThesis2010}.

%%%%%%%%%%%%%%%%%%%%%%%%%%%%%%%%%%%%%%%%%%%%%%%%%%%%%%%%%%%%%%%%%%%%%%%%%
\subsection{Lifetimes per segment}
\label{SIsec:seg_corr}
Module name: \py{segmented\_crosscorrs.py}\\
Input files: \py{timestamps\_chX\_bin.parquet},\py{segmentbounds.txt}\\
Output files: \py{corr\_seg.txt}, \py{segment\_fit\_data.csv} \\

This script again imports  the photon events  from the parquet files, as well as the segment bounds (jumptimes) calculated and exported  by \py{CPA\_wrapper} \ref{SIsec:segmenting}. Looping through the segments, the events on either detector are correlated with the events in the reference channels using \py{correlate\_jit},  following the method of~\cite{Wahl2003}. The correlations are calculated over a time range equal to  the (assumed) repetition rate of the excitation laser. This gives two $2\times Q-1 \times q$ matrix, with $Q$ the number of found CPA jumps (including time series start and stop) and $q$ the ratio of $\tau_L$ the time between excitation pulses over the timing card resolution. Thus each each of the $2(Q-1)$ (factor 2 for the two detectors) rows represents a fluorescence decay histogram for a segment. Since the correlation calculations can be quite lengthy, they are saved to corr\_seg.txt, for loading as needed. In addition, we speed up these calculations significantly by making use of Numba, which is a so-called just-in-time (JIT) compiler. This allows a specified section of Python code to be accelerated by just in time compilation. For these segmented correlations, this gives a speedup of between one and two orders of magnitude. This JIT compiler is applied for all correlations calculated, others including the $g^2(\tau)$, the full decay trace, and the coarsened autocorrelation.

Following the interpretation-agnostic conversion from timestamps to decay histograms per segment,  an   exponential decay (by default single exponential --- several other dependencies are preprogrammed and could be chosen by the user)   is fitted to each segment. This is done using the maximum likelihood estimator (MLE) relevant to Poisson statistics, explained by Bajzer et al.~\cite{Bajzer1991} through the routine \py{fitexp\_MLE}. 
Before fitting, the data is  prepped. This includes timeshifting the photon channels such that the peak of the decay histograms lands at $\tau=0$ (each might have a different shift, determined by the routine as \py{Tau0A\_bin} and \py{Tau0B\_bin}),  reversing the time axis (exponential decay tail towards positive time, undoing reverse start stop) clipping out negative times (detector rise), and includes an option for clipping out user-specified data intervals.  This is useful for experimental data, where electronic ringing in DPC/TCSPC chips sometimes causes artefacts that should not be fitted. Note that excluding this range from the fit is \textit{not} the same as setting the counts for this range to zero, instead removing them entirely.

After the data prepping per detector channel, the detector channels are spliced, so that decay rates are determined from the full photon record (sum of both detectors). Importantly for MLE-fitting of Poisson distributed data in which the probability distribution is linked to absolute count values, the data is neither scaled, nor background substracted.  By fitting a single exponential $A\exp(-i\gamma t) +b$, three values and their respective errors are obtained. The routine exports $A$, $\gamma$, the error in $\gamma$, and $b$. It should be noted that in the subsequent processing routines only $\gamma$ and $\delta \gamma$ are used, as the best estimate for the count rate per segment simply follows from the number of counts in a segment $N_q$ divided by the segment duration $T_q$ (error follows from the square root of the number of counts, divided by the segment duration).
The  fitted values are  saved, and can be reloaded as a pandas dataframe in \py{segment\_fit\_data.csv}.  The fit routine starts from an initial guess for parameter values that are estimated from the data itself.

For insightful explanations of the maximum likelihood estimator, we refer the reader to~\cite{FrimmerThesis2012, Bajzer1991}.

%%%%%%%%%%%%%%%%%%%%%%%%%%%%%%%%%%%%%%%%%%%%%%%%%%%%%%%%%%%%%%%%%%%%%%%%%
\subsection{Grouping, determining most likely levels}
\label{SIsec:grouping}
Module names: \py{grouping\_wrapper.py}\\
Input files: \py{segmentbounds.txt}\\
Output files: \py{grouping.txt}\\

This routine operates on the result of \py{CPA\_wrapper}, i.e., the file \py{segmentbounds.txt} produced by it. This consists of a list of jump times ($t_1, t_2,..$) and (cumulative) counts per segment,  which  convert into segment durations $T_q=t_q-t_{q-1}$, counts per segment $N_q$ and  per-segment count rates $I_q=N_q/T_q$. 
Following the main text, the segments are clustered into a first guess for the actual Bayesian grouping  analysis using a recursive hierarchical clustering algorithm that crawls through a data set of $Q-1$ segments in $Q-1$ steps,  in each step grouping the segments with the nearest intensity together.  We find that the final result of the Bayesian grouping analysis is not very dependent on this initial guess, which allows for a shortcut. For this reason we us a faster version of the original hierarchical clustering proposed by~\cite{Watkins2005} for large data sets, which consists of hierarchically grouping small blocks of 300 subsequent segments to 30 levels each, concatenating those, and   then grouping.  The result is  groupings of the data in a range of $\nG$ levels (in our analysis we didn't go beyond $(\nG)=20$). 

After this initial guess, the segment allocation to intensity levels, and the level-values are optimized following Section 2.3.2 of~\cite{Watkins2005}. Here, the segments, divided into exactly $\nG$ groups by the initial hierarchical clustering, are iteratively swapped out, until the most likely grouping on the assumption of exactly $\nG$ levels is achieved, with the best estimate for the intensity levels.
Thus for each  hypothetical distribution in $\nG\in\py{mingroups}, ..., \py{maxgroups}$ levels, the most likely assignment of each segment to an intensity level $\mathcal{I}_1, ... \mathcal{I}_{\nG}$ is determined. This gives a ``trajectory'' of the emitter through the different levels, and is the which-level information for each segment, for each of the tested $\nG$. Finally, to determine which value of $\nG$ most likely describes the data, the Bayesian Information Criterion (BIC) is calculated as function of  $\nG$. This gives the likelihood of a given $\nG$ to be the true, underlying set of levels. For simulated $m_0$-level dots indeed, the BIC peaks at $\nG=m_0$

The grouping routine returns its result as plots and  ASCII data in \py{grouping.txt}. This data folder contains three sets of information. 
The first is a list of the values of the Bayesian Information Criterion, discussed in the main text and shown in Fig. 4a. 
The second is a square matrix, where the values $b_{\nG,m}$ describe the fraction of time spent in state $m$, given that the grouping algoithm had  $\nG$ intensity levels available to it. This yields figures such as Fig.4b in the main text. 
Lastly, the file contains a 2D matrix that describes the ``trajectory'' of the emitter through the available levels, given that $\nG$ levels are available to it. There is no figure of this in the main text, though Fig. 1D. provides a sketch.

%%%%%%%%%%%%%%%%%%%%%%%%%%%%%%%%%%%%%%%%%%%%%%%%%%%%%%%%%%%%%%%%%%%%%%%%%%%%%%%%%
%%%%%%%%%%%%%%%%%%%%%%%%%%%%%%%%%%%%%%%%%%%%%%%%%%%%%%%%%%%%%%%%%%%%%%%%%%%%%%%%%

%%%%%%%%%%%%%%%%%%%%%%%%%%%%%%%%%%%%%%%%%%%%%%%%%%%%%%%%%%%%%%%%%%%%%%%%%%%%%%%%%
\subsection{Postprocessing segmented data, plot routines}
\label{SIsubsec:traceplot}
\subsubsection{Time trace}
\label{SIsec:trace}
Module names: \py{plot\_timetrace.py}\\
Input files: \py{timestamps\_chX\_bin.parquet}, \py{segmentbounds.txt}, \py{segment\_fit\_data.csv}\\
Output files: \py{none}\\

This script functions purely as a visualization of the behavior of the emitter as a function of time. For plotting purposes the photon events are binned in equidistant time bins to obtain an intensity trace in time, which is plotted alongside the   intensities and decay rates retrieved by CPA (i.e., without any binning). Additionally, a histogram of the intensity levels in the time trace is plotted. This is an unweighted histogram, where all CPA-determined segments  have equal weight, independent of their duration. The routine also plots a time trace  of the fitted decay rates  with its corresponding histogram besides it.  For plotting purposes, the routine attempts to guess reasonable plot axes from the data.

%%%%%%%%%%%%%%%%%%%%%%%%%%%%%%%%%%%%%%%%%%%%%%%%%%%%%%%%%%%%%%%%%%%%%%%%%%%%%%%%%
\subsubsection{Switching time statistics}\label{SIsubsec:switching}
Module names: \py{switchingtimehistogram.py}\\
Input files: \py{segmentbounds.txt}\\
Output files: \py{switchinghist.txt}\\

In this section of the code, the time durations $T_q$ of all found CPA segments are taken into a histogram with logarithmic bin, implementing apropriate normalization to convert the segment durations to a segment duration probability distribution function distribution. Next, the routine fits a power law, as typically observed for  many single emitters follow in their on/off times. Because CPA is more likely to miss switching events if they are very short (segment with $< 50$ to $100$ photons), the found histogram will slope off from the initial power law at short switching times.  The routine will therefore fit a power law to the part of the curve at larger switching times than a chosen threshold value. It uses the minimize module from \py{scipy.optimize} to do a least-squares fit to the data, reporting as output the fitted power law exponent and a plot to assess the quality of fit.

%%%%%%%%%%%%%%%%%%%%%%%%%%%%%%%%%%%%%%%%%%%%%%%%%%%%%%%%%%%%%%%
\subsubsection{FDID and FLID diagrams}\label{SIsubsec:FDID}
Module names: \py{FDID\_wrapper.py}\\
Input files: \py{segmentbounds.txt, segment\_fit\_data.csv}\\
Output files: \py{FDID.txt}, \py{FLID.txt}\\

In this script, the segment intensities and segement-resolved decay rates are compiled into fluorescence \textit{decay rate} intensity diagrams (FDIDs) and/or fluorescence \textit{lifetime} intensity diagrams (FLIDs). Either of these diagrams can be made by setting a boolean in \py{EntireWorkflowScript.py}. For these processes, the $I_q, \gamma_q$ values of all segments exported into \py{segment\_fit\_data.csv} from \py{CPA\_wrapper.py} are used to build the 2D histograms. The process is nearly identical for either diagram, after decay rates and the error thereof are converted into lifetimes, and both options are contained in the same routine. 
For the FDID and/or FLID, instead of simply compiling a histogram of square bins, each segment contributes a smooth $2D$ Gaussian, the widths of which are determined by the \textit{error} on the segment decay rate and the estimated uncertainty in the count rate for that segment, derived from the fact that counts per segment are Poisson distributed.  The weight of each Gaussian is defined as its integral. As explained in the main text, three different definitions of  weight per segment  can be used: (1) all CPA segments have the same weight, (2) the weight of a segment is determined by the number of photon events in it, (3) the segment weight is determined by the time duration of the segment. When applying weighting, it is the volume under each 2D Gaussian that is weighted.  A single routine \py{FDID\_functions} constructs the FDIDs and FLIDs as sum of Gaussians, with the desired choice of weight passed as argument from the routine \py{FDID\_wrapper} that handles file input, and plotting. For plotting purposes, the wrapper routine attempts to guess reasonable plot axes from the data. The resulting FDIDs and FLIDs are output as ASCII files and plots.
\vspace{0.3cm}Output:
\begin{itemize}
    \item FDID\_prio\_none and/or FLID\_prio\_none, 2D array, the FDID/FLID where all CPA segments have the same weight
    \item FDID\_prio\_cts and/or FLID\_prio\_cts, 2D array, the FDID/FLID where the CPA segments are weighted by the amount of counts
    \item FDID\_prio\_dur and/or FLID\_prio\_dur, 2D array, the FDID/FLID where the CPA segments are weighted by their duration
    \item FDIDextent and/or FLIDextent, tuple, the (x1, x2, y1, y2) extent of the FDID/FLID. Used for plotting
\end{itemize}

%%%%%%%%%%%%%%%%%%%%%%%%%%%%%%%%%%%%%%%%%%%%%%%%%%%%%%%%%%%%%%%%%%%%%%%%%%%%%%%%%
\subsubsection{Memory effects}\label{SIsubsec:memoryeffects}
Module names: \py{FLID\_wrapper.py}\\
Input files: \py{segmentbounds.txt, segment\_fit\_data.csv}\\
Output files: \py{none}\\

This script converts the segmentation results from \py{CPA\_wrapper} into correlative plots that highlight if there are memory effects and correlations between segment observables. The routine reads in \py{segment\_fit\_data.csv}, i.e., sequencies of intensities  $I_q$, the fitted decay rates $\gamma_q$, and segment durations $T_q$. Due to the large range of segment durations, plots are based on  $\log_{10}T_q$.  The generated plots report on:

\begin{enumerate}
    \item \textbf{Aging:} Time trace of $I$, $\gamma$ and $\log_{10}T_q$ versus `wall clock time'. In these figures, the aging of an emitter can be observed, e.g. if the dot becomes dimmer over time, or segment lengths grow or shrink.
    \item \textbf{Concurrent observable correlations}  A set of 2D histograms of the $\log_{10}T_q$ versus  (1) intensity $I_q$ and (2) the fitted decay rate $\gamma_q$. This can be used to detect if there are correlations between these quantities. These diagrams are anaologous to FDID or FLID diagrams, which are just 2D histograms of $I_q$ versus $\gamma_q$.
    \item  \textbf{Memory} A set of $2 \times 3$ 2D histograms, which plot conditional probabilities of $x_{n}$ against $x_{n-1}$ and $x_{n_2}$, with $x \in I,\,\gamma,\,\log_{10}T_q$. These are 2D histograms that examine the same observable along both their axes, but at subsequent steps in the segment history. For instance, plotting $\gamma_{n}$ against   $\gamma_{n-1}$ highlights if subsequent decay rates are uncorrelated, or correlated.
    \item \textbf{Segment sequence autocorrelations}  of the lists $I_q,\,\gamma_q$ and $\log_{10}T_q$.  Note that autocorrelating a list such as $I_q$ is not equivalent to calculating the intensity autocorrelation, i.e., autocorrelating $I(t)$ as provided by \py{Houel\_autocorr.py}, since the segment intensity list $I_q$ contains intensity levels of segments with highly unequal duration, whereas the autocorrelation of $I(t)$ depends on the switching dynamics in time. The segment sequence autocorrelation brings out the number of switching cycles over which the correlated observable shows a memory.
\end{enumerate}

%%%%%%%%%%%%%%%%%%%%%%%%%%%%%%%%%%%%%%%%%%%%%%%%%%%%%%%%%%%%%%%%%%%%%%%%%%%%%%%%%
%%%%%%%%%%%%%%%%%%%%%%%%%%%%%%%%%%%%%%%%%%%%%%%%%%%%%%%%%%%%%%%%%%%%%%%%%%%%%%%%%

\subsection{Unsegmented data analysis}
%%%%%%%%%%%%%%%%%%%%%%%%%%%%%%%%%%%%%%%%%%%%%%%%%%%%%%%%%%%%%%%%%%%%%%%%%%%%%%%%%
\subsubsection{Decay trace and fit, entire dataset}\label{SIsubsec:totaldecay}
Module names: \py{unsegmented\_crosscorrs.py}\\
Input files: \py{timestamps\_chX\_bin.parquet}\\
Output files: \py{corr\_unseg.txt}\\

The entire data set is processed by \py{correlate\_jit}  to obtain the correlation between detector channel and reference channel, leading to two decay traces (for detector A, and B). As in the case of the segmented decay trace determination routine, this data corrected by zeroeing the time axis, reversing the start-stop reversed time axis and removing the detector artefacts (risetime, electronic ringing).  The two detector channels are then merged to obtain a fluorescence decay histogram. To this, an $n$-exponential curve can be fitted. The same Poisson-MLE is used as  for the unsegmented data. Using the Minimize function from the Scipy Optimize module, we minimize our merit function such that a n-exponential decay 
\begin{equation*}
    F_{{A_j}, {\gamma_j}, B}(\tau) = b + \sum_{j=1}^n A_j \exp(\gamma_j \tau)
\end{equation*}

is fitted to the calculated decay trace. Here, $A_j$ and $\gamma_j$ are the proportionality constants and the decay rates of the decay $j$, respectively, and $b$ describes constant background noise. When fitting, we optimize for ${A_j, \gamma_j, B}$. How the fit merit function is calculated from the data and the fit function is discussed in the main text.

%%%%%%%%%%%%%%%%%%%%%%%%%%%%%%%%%%%%%%%%%%%%%%%%%%%%%%%%%%%%%%%%%%%%%%%%%%%%%%%%%
\subsubsection{Autocorrelation $g^{(2)}(\tau)$}\label{SIsubsec:autocorr}
Module names: \py{g2\_tau.py}\\
Input files: \py{timestamps\_chX\_bin.parquet}\\
Output files: \py{g2.txt}\\

This function reads the photon detection time stamps from parquet file, and then calculates the unnormalized intensity autocorrelation $g^{(2)}(\tau)$ by correlating the two detector channels following the method outlined in~\cite{Wahl2003} to obtain $g^{(2)}(\tau)$. For data taken in a Hanbury-Brown Twiss setup, this should reveal antibunching, as is the case for the example experimental data set. The simulated data generated by running the provided routine also antibunches. It should be noted that the routine does not correct for relative constant offsets in channel timings due to, e.g., differing electronic delays (cable lengths), which, if present, expresses as a small shift of the the $g^{(2)}(\tau)$  along the the time-axis.

\subsubsection{Long-time intensity correlations}\label{SIsubsec:longcorr}
Module name: \py{Houel\_autocorr.py}\\
Input files: \py{timestamps\_chX\_bin.parquet}\\
Output files: \py{Houel\_autocorr.txt}\\

Houel et al.~\cite{Houel2015} proposed that intensity autocorrelations on the time scales of milliseconds to seconds express an unbiased assessment of intermittency characteristics, relating at least for binary on/off behaviors to power law exponents of dwell times.  While Houel et al. operated on intrinsically timebinned data (camera frames), we provide a routine that operates directly on timestamps.  The routine \py{Houel\_autocorr.py} reads in timestamps for two detector channels from Parquet files,  splices the two channels in a single array of time stamps, and then calls \py{correlate\_jit} to autocorrelate the timestamp array. In view of the large time span over which the correlation is requested compared to the time-discretization intrinsic to the data (second versus 0.1 ns), a standard correlation as used for $g^{(2)}(\tau)$ is prohibitively slow. Therefore \py{correlate\_jit} provides a version of the timestamp correlation algorithm with logarithmically increasing steps in $\tau$ through a data coarsening approach as outlined in~\cite{Wahl2003}. As opposed to the unnormalized $g(2)$ for antibunching, now the output that we refer to as ACF is normalized. To ACF-1 a  powerlaw with exponential roll off
\begin{equation*}
At^{-C}\exp(-Bt)
\end{equation*}
is fitted where we use the minimize function of Scipy's Optimize module and a least-squares fit. According to Houel et al.~\cite{Houel2015} the fit parameters trace back to power law exponents in the case of binary (on/off) intermittency. 

\subsection{Auxiliary routines}
The role of the remaining routines in the processing toolbox is
\begin{itemize}
    \item \py{acc\_functions.py} contains shorthand for auxiliary mathematical functions, and 2D plot color map
    \item \py{EntireWorkflowScript.py} example script executing full sequence of processing steps. Booleans in the file set the requested tasks (e.g., simulated or experimental data), but otherwise no data is passed from function to function. The only dependency is through the intermediate data files
    \item \py{loaddata\_functions.py} contains file input output commands for reading and writing the various files in the appropriate formats and directories
    \item py{preamble\_measured.py} contains the file paths to the example experimental data, and required output data, as well as metadata required for processing. These include the TCSPC discretization bin size of the experiment (0.165 ns, Becker and Hickl DPC-230) and the laser repetition rate (10 MHz) in the experiment that generated the data.
    \item py{preamble\_simulated.py} contains the file paths to simulated emitter data, and required output data, as well as metadata required for processing. It should be noted that parameters required for processing, i.e. the assumed time bin discretization unit and laser repetition rate need to be specified in the file, and need to be consistent with how the data was generated. They are not automatically passed from the simulation routine to maximize the direct exchangeability of experimental and simulated data for testing the toolbox.
\end{itemize}

\subsection{Binning instead of using CPA}
In the sections above, we have discussed the analysis following segmentation of the data using CPA. However, the same analysis can be applied following segmentation using equidistant bins. If the user is interesting in applying equidistant binning, this can be done by flipping the Boolean \py{CPA\_insteadof\_binned} in the workflow module. The output from this segmentation and analysis will be saved in separate folders, to avoid overwriting CPA-sectioned data. 
The only script that will not work when equidistant binning is applied, is of course \py{switchingtimehistogram}, as its function starts off with building a histogram of the duration of the segments.

%%%%%%%%%%%%%%%%%%%%%%%%%%%%%%%%%%%%%%%%%%%%%%%%%%%%%%%%%%%%%%%%%%%%%%%%%%%%%%%%%
%%%%%%%%%%%%%%%%%%%%%%%%%%%%%%%%%%%%%%%%%%%%%%%%%%%%%%%%%%%%%%%%%%%%%%%%%%%%%%%%%

\section{Simulate N-level emitters}\label{SIsec:simulate_dot}

Module name: \py{simulate\_qdot.py}\\
Output files: \py{timestamps\_chX\_bin.parquet, nominalsegmentation.csv}\\

We provide a routine to generate  time-tagged data of an $m_0$-level single emitter for Monte Carlo simulations testing the toolbox, or for testing how assumed photophysics would appear in experiments. This routine is provided in a separate folder from, and is in terms of code independent from, the processing toolbox. It produces timestamps in parquet file format that, together with an appropriately specified \py{preamble\_simulated.py} file can be processed in exactly the same manner as the example experiment data set . 

In this section, we will describe the method used to simulate time-tagged data of an $m_0$-level single emitter with MC simulations. The entire workflow of generating a dot can be executed by running  \py{entire\_simulate\_qot\_workflow.py}. This script generates photon time stamps, and also creates plots and a reference datafile to examine the assumed and simulated intermittency behavior.

The simulation of the emitter behavior happens in a number of steps, as follows:
\subsection{Drawing the segmentation information}
First we randomly draw the random segmentation sequence according to which the photon time stamps will be subsequently randomly drawn, using the following procedure:
\begin{enumerate}
\item  The user specifies the number of levels \py{nlevels}$=m_0$, and the associated nominal intensity levels, decay rates, and dwell-time power law exponents through the preamble file.
\item  A random list is created with indices that label the order in which the  intensity levels are visited. For each entry in the list, the index is drawn independently and all levels have equal likelihood. If an entry is a direct repeat of its predecessor, it is removed from the list. The segment lengths are drawn using the following formula~\cite{Clauset2009}
\begin{equation}
    s = (a^\beta +(b^\beta - a)x)^{1/\beta}
\end{equation}
where $a$ is the short time cut off \py{Tshortest\_ns}, $b$ is the long time cut off\py{Tlongest\_ns}, and $\beta$ is $1-\alpha$, with $alpha$ the specified power law exponent for the given level. $x$ is a random number generated with numpy's random.rand module, and has units of dtau\_ns.
\end{enumerate}

\begin{figure}[t]
  \includegraphics[width=0.9\columnwidth]{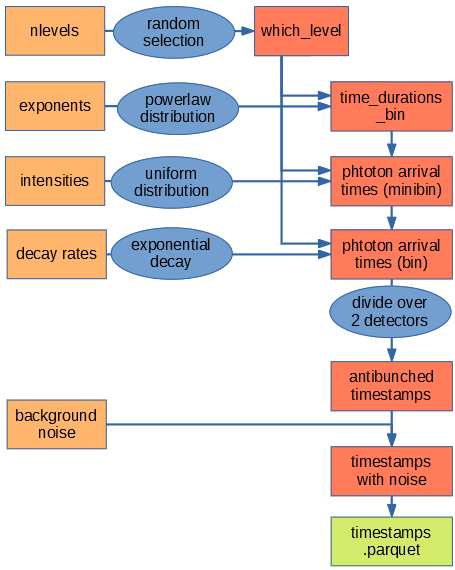}
  \caption{Flow chart of the routine to simulate multi-level emitters.}
\label{fig:SI_simulatedot}
\end{figure}

\subsection{Drawing photon events}
The next step is to draw photon timestamps according to the randomly drawn segmentation sequence. The segmentation sequence specifies a list of jump times $t_1, t_2, ..., t_Q$ and a concomitant sequence of levels visited, which specifies the nominal count rate, and decay. We draw the photon events  in a two step process:
\begin{enumerate}
    \item Since we envision simulating a single emitter in a pulsed experiment, every laser repetition cycle there can be at most one photon resulting from an excitation event.  That means that for a given nominal count rate there is a known probability $p$ that one photon is generated. We thus draw photon time stamps with a resolution equal to the laser repetition time $\tau_L$ by separating a segment of length $T_q$ in $T_q/\tau_L$ bins, drawing for each bin a random uniformly distributed number between $0$ and $1$, and conclude that a given cycle corresponds to a photon if the corresponding drawn random number exceeds $p$. 
    In the provided example, $\tau_L=100$ ns, which gives the granularity with which the time stamps are defined at this step. The probability of a photon event is given by $p=I_q\times\tau_L$.  It should be noted that drawing rare events according to uniform distribution results in Poisson distributed count rates upon binning. Programmatically, only the list of timestamps labelling laser cycles with a resulting emitted photon are kept, and stored in list \py{timestamps\_chR\_bin}.  This ultimately is the reference channel for the simulated two-detector TCSPC measurement. 
    \item Next we fine-grain the photon time stamps, to account for the fluorescent decay trace. For each timestamp in the reference channel, a photon arrival time relative to the laser pulse is drawn through numpy's random.exponential module,  with the appropriate decay rate from \py{g\_lst} for the relevant segment. 
    \item from a memory management viewpoint, step 1 is prohibitive if applied to the entire desired time trace (typically $10^{11}$ laser cycles, for only 
    $10^7$ photon events), so we apply it segment by segment. To avoid the slow python append function, the timestamps are stored in a preallocated, overdimensioned container array, which is clipped to size after generating all the time stamps.  Since we provide a simulation toolbox to mimic TCSPCS experiments, all times are expressed as integer multiples of an assumed card binsize \py{dtau\_ns}. This allows us to express all time quantities in int64 format.
    \item uncorrelated dark noise for two detector channels is drawn according to step 1, with in step 2 a uniform random arrival time distribution.
\end{enumerate}

\subsection{Distributing events over channels}
The outcome of the previous step is a set of time stamps \py{timestamps\_chR\_bin} corresponding to laser pulses with a concomitant detection, and a list of delay times relative to the laser pulses. These are translated into antibunching detector channels as follows:
\begin{enumerate}
    \item We construct photon arrival time stamps in reverse start stop, by taking \py{timestamps\_chR\_bin} adding the delay times, and subtracting the duration of exactly one laser period. This leads to a list of photon time stamps \py{timestamps\_bin}.
    \item For each event in \py{timestamps\_bin} we randomly assign it to detector channel A or detector channel B, with equal and independently drawn probability.
    \item Finally we splice the generated uncorrelated detector noise events  \py{noiseclicks\_chX} with $X \in A,B$ into the data stream. The concomitant laser pulse arrival times  was previously sent into \py{timestamps\_chR\_bin}. 
\end{enumerate} These operations ensure that the two simulated detector channels will show antibunching in $g^{(2)}(\tau)$, with a visibility limited by the signal to background count rate. The background noise photon events are only added now, as adding them earlier would cause the noise to antibunch as well.

\subsection{Input and output files}
The simulation module requires input parameters, which are specified through the module \py{preamble\_for\_simulation.py}. The input parameters include parameters of the simulated emitter (number of levels, concomitant nominal count rates, decay rates, and power law exponents)  and of the simulated measurement (laser repetition time, time discretization bin size,  background count rate, and total wall clock length of the measurement record). 

It is important to note that most of the variables are independent of the variables used for processing the data, \textit{except} the variables that pertain to the properties of the ``measurement'' itself. These are what in a measurement corresponds to the timing resolution of the counter card and the repetition rate of the excitation, called \py{dtau\_ns} and \py{minitimebin\_ns} respectively. When analyzing simulated data, the values of these two variables must be the same as in data generation.

As output three lists of timestamps, with timestamps as 64 bit integers in units of the timing resolution \py{dtau\_ns}, representing two detectors in a Hanbury-Brown Twiss configuration, and a laser pulse reference channel for both. These are saved in a compact binary format as  parquet files. 
As a service to the user, the simulation writes out the datafile   \py{nominalsegmentation.csv} with the jump times, and the concomitant sequence of nominal segment count rates and decay rates according to which the data was drawn. This can e.g. be used to check the accuracy of the CPA, grouping,  and lifetime fitting routines.
\\

\section{Example experimental data}
We provide select example data sets with the toolbox that were  obtained on a  CsPbBr$_3$ quantum dots, as reported in Ref.~\cite{Palstra2021expt} (dots 2, 4, 6, 8, 16, with reference to supplement of that paper). While the full experimental protocol is reported in that paper, here we report the information pertinent for interpreting the time tagged data. The data was obtained using pulsed excitation at 10 MHz repetition rate (LDH-P-C-450B pulsed laser diode, PicoQuant) at 10 MHz, with confocal excitation and collection through an oil objective (Nikon Plan APO VC, NA=1.4). Detection was on two fiber-coupled avalanche photodiodes (APDs) (SPCM-AQRH-14, Excelitas) in a Hanbury-Brown \& Twiss configuration. The APDS were coupled to a photon correlator (Becker \& Hickl DPC-230). It uses a 0.165 nm temporal binsize, and simultaneously records absolute photon arrival times from both APDs, and the arrival times of laserpulses subsequent to photon detection on any detector.  We note that this means our decay traces are taken  in reverse start-stop configuration, as is standard in TCSPC to reduce data rates. A small time interval centered at around 30 ns is subject to an electronic artefact which we attribute to a ringing in the DPC-230 TDS timing chips. This interval is included in the supplied data - the toolbox is set up to  exclude it from the analysis. Our measurements are taken over 120 seconds of acquisition time.   The timetagged data was converted from the Becker and Hickl dataformat using C$\#$ code that uses the Becker and Hickl SPCM DLL toolkit. We note that the data of any of the TCPSC/timetagging equipment vendors can be analyzed by our toolbox, provided it is converted from proprietary format to timetagged channel data, saving the timetag lists expressed in units of the card bin size as ordered 64-bit integer arrays, with a single parquet file per channel.
 
\section{Dependencies}
The code is posted on GitHub  \py{https://github.com/AMOLFResonantNanophotonics/CPA/}. This code has been developed in Python 3.8. It requires to install the libraries \py{numpy} [1.19.2],  \py{matplotlib} [3.3.2], \py{pandas} [1.1.3], as well as \py{numba} [0.15.2]  for jit-speedup of the Wahl algorithm, and  \py{pyarrow} [2.0.0]\ for dealing with parquet files. Numbers in square brackets [.] indicate the version installed when developing the code.

\section{Archived versions}
The first version of the code, and the example data sets are archived via \url{DOI: 10.5281/zenodo.4557226}

%%%%%%%%%%%%%%%%%%%%%%%%%%%%%%%%%%%%%%%%%%%%%%%%%%%
%\bibliographystyle{Style/vancouver}
%\bibliography{library}

%\bibliographystyle{apsrev}
%\bibliography{PalstraPython}
%\end{document}
%%%%%%%%%%%%%%%%%%%%%%%%%%%%%%%%%%%%%%%%%%%%%%%%%%%

\end{document}